\documentclass[aps,pra,reprint,showkeys,longbibliography]{revtex4-2} 
\usepackage{amsmath}
\usepackage{amssymb}
\usepackage{graphicx} 
\usepackage{bm} 
\usepackage{xcolor}
\usepackage[toc,page]{appendix}
\usepackage{braket}
\usepackage{mathrsfs}

\makeatletter

\@ifclasswith{revtex4-2}{draft}{%
    \usepackage[notref,notcite]{showkeys}
    
}{}

\makeatother

\newcommand{\nn}{\nonumber}

\begin{document}

\title{Classical probabilistic realisation of quantum double-slit interference}

\author{Christof Wetterich}
\affiliation{Institut f\"ur Theoretische Physik\\
    Universit\"at Heidelberg\\
    Philosophenweg 16, D-69120 Heidelberg}

\begin{abstract}
We demonstrate how the interference effects for a quantum particle in the double-slit experiment 
can be described by classical probabilities. We investigate a classical field theory
for a complex scalar field with probabilistic initial conditions. 
A central element are conserved charges leading to the concept of particles.
These are statistical observables which describe properties of the probability distribution for field configurations. 
The conserved charges define subsystems for particle excitations of a vacuum state. 
The classical probability distribution for the one-particle subsystem can be expressed in terms of a complex wave function. 
The Liouville equation for the classical probability distribution implies that the time evolution of this wave function obeys the Schr\"odinger equation for a quantum particle in a potential. 
The potential arises from space-dependent external fields in the otherwise relativistic classical  field theory. 
It can be chosen arbitrarily, realizing the typical quantum effects of interference, tunneling or discrete energy spectra.
\end{abstract}

\maketitle


\indent The double-slit experiment is a corner stone for the particle-wave duality in quantum physics. 
The interference demonstrates the wave aspect, while individual clicks in detectors underline the 
particle aspect. It has been advocated that quantum mechanics can be embedded in a classical 
probabilistic setting \cite{CWO2, CWPO, CWEQMCS}. This claim should show that the outcome of the double-slit experiment can be 
realized in a system described by a classical probability distribution. This demonstration is the purpose 
of the present note. We construct classical probability distributions which implement the dynamics of 
a quantum particle in an arbitrary potential. The setting for the double-slit experiment corresponds
then to a suitable potential, as for quantum mechanics.

\indent Our findings can be summarized as follows: We consider a classical complex scalar field $\sigma(t,\vec x)$ which obeys the field equation
\begin{equation}
\partial_t^2 \sigma
=
-B\sigma\,,
\quad
B=-\Delta+\bigl(m+V(\vec x)\bigr)^2\,,
\label{eq:I1}
\end{equation}
with Laplace operator $\Delta=\partial_x^2=\sum_k \left( \partial/\partial x_k\right)^2$.
The potential $V(\vec x)$ can be seen as an external or background scalar field, somewhat similar to electrostatics.
At every $t$ we describe the classical probabilistic system by a probability distribution for field configurations $w(t;\sigma (\vec{x}),\pi (\vec{x}))$, with $\pi (\vec{x}) = \partial_t \sigma (\vec{x})$.
Its time evolution is given by the classical statistical Liouville equation.
We construct particular "one-particle probability distributions" $w^{(1)} (t)=\bigl(q^{(1)}(t)\bigr)^2$, where
\begin{equation}
q^{(1)} (t)
=
\int_{\vec x,\vec y}
\psi_S(t,\vec x)\psi_S^*(t,\vec y)\,
C(\vec x,\vec y)\,
q^{(0)}
-
q^{(0)}\,,
\label{eq:I2}
\end{equation}
with 
\begin{equation}
\label{eq:I3}
C(\vec x,\vec y)
=
\bigl(\tilde\sigma^*(\vec x)-i\tilde\pi^*(\vec x)\bigr) 
\bigl(\tilde\sigma(\vec y)+i\tilde\pi(\vec y)\bigr)\,,
\end{equation}
and
\begin{equation}
q^{(0)}
=
\mathcal{N}_0
\exp\left\{
-\int_{\vec x}
\left(
\tilde\sigma^*(\vec x)\,\tilde\sigma(\vec x)
+
\tilde\pi^*(\vec x)\,\tilde\pi(\vec x)
\right)
\right\}\,.
\label{eq:I4}
\end{equation}
Here we define
\begin{equation}
    \label{eq:4A}
    \tilde \sigma (\vec x) = \sqrt{2} H_S^{1/2} \sigma (\vec x)\,, \quad
    \tilde \pi (\vec x) = \sqrt{2} H_S^{-1/2} \pi (\vec{x})\,.
\end{equation}
The hermitian Hamiltonian $H_S$ reads
\begin{equation}
H_S
=
\sqrt{B}
\approx
-\frac{\Delta}{2m}+V(\vec x) + m\,,
\label{eq:I5}
\end{equation}
where the second expression employs the "non-relativistic limit" $-\Delta \ll m^2$; $|V(\vec{x})|\ll m$, and the constant $m$ does not matter.
The family of one-particle probability distributions is parametrized by the complex "Schr\"odinger wave function" $\psi_S(t,\vec x)$. 
It is normalized if $q^{(0)}$ is normalized by a suitable choice of $\mathcal{N}_0$.

\indent Our central result states that the Liouville equation for $w^{(1)}$ implies that $\psi_S$ obeys the
Schr\"odinger equation for a quantum particle in a potential $V(\vec x)$.
In turn, every solution of the Schr\"odinger equation for $\psi_S$ constitutes a solution of the Liouville equation for $w$. 
Thus the family of one-particle probability distributions $w^{(1)}(t)$ describes a closed subsystem in the space of probability
distributions, which shows all dynamical features of a quantum particle in an arbitrary potential $V(\vec x)$. 
For suitable potentials this subsystem realizes the characteristic features of quantum mechanics,
as interference in the double slit experiment, tunnelling through potential barriers, or the discrete spectrum of the quantum energy in the Coulomb potential.
The observables of position, momentum or angular momentum of the quantum particle obtain from statistical observables of the classical field theory by projection on one-particle states. 
Their expectation values are given by the quantum rule without invoking additional ``quantum axioms''.

\indent More in detail, the one-particle probability distribution $w^{(1)} [\rho_S]$ is a linear
functional of the quantum density matrix
\begin{equation}
\rho_S(\vec x,\vec y)
=
\psi_S(\vec x)\psi_S^*(\vec y)\,.
\label{eq:I6}
\end{equation}
Its time evolution according to the Liouville equation obeys
\begin{equation}
\partial_t w^{(1)}[\rho_S]
=
-\hat L\,w^{(1)}[\rho_S]
=
w^{(1)}[\partial_t\rho_S]\,,
\label{eq:I7}
\end{equation}
where the Liouville operator $\hat L$ reads
\begin{equation}
\hat L
=
\int_{\vec x}
\left\{
\pi\,\frac{\partial}{\partial\sigma}
+
\pi^*\,\frac{\partial}{\partial\sigma^*}
-
B\left(
\sigma\frac{\partial}{\partial\pi}
+
\sigma^*\frac{\partial}{\partial\pi^*}
\right)
\right\}\,,
\label{eq:I8}
\end{equation}
Here $\partial_t \rho_S$ is given by the von-Neumann equation with Hamiltonian $H_S$,
\begin{equation}
\partial_t\rho_S(\vec x,\vec y)
=
-i\left(
H_S(\vec x)\rho_S(\vec x,\vec y)
-
\rho_S(\vec x,\vec y)H_S(\vec y)
\right)\,.
\label{eq:I9}
\end{equation}
Eq.~\eqref{eq:I7} can be verified by direct computation for general quantum density
matrices beyond the pure state case \eqref{eq:I6}.

\indent The structure behind this result is based on conserved charges and
associated particle numbers. They play a double role. 
First, the particle numbers are integer, which accounts for the discreteness of
possible measurement values for certain observables. 
Second, the particle numbers are conserved.
This allows for the definition of subsystems within the space of probability distributions
which are closed under the time evolution -- namely subsystems with a fixed particle number.
One finds both particles and antiparticles with opposite charge.
Particles can be viewed as excitations of a vacuum state which is given by $q^{(0)}$ in eq.~\eqref{eq:I4}.

\indent All these concepts are familiar from quantum field theory.
We show that the quantum field theory for a complex scalar field is actually a subsystem of the more general probabilistic classical field theory.
Probabilistic classical field theories of the type discussed here have been investigated as a ``classical approximation'' to a quantum field theory \cite{STA, BER}. 
This is not our approach. 
We start from the classical field theory as the basis, and show that it contains a quantum field theory as a subsystem. 
The ``quantum'' emerges from the ``classical''.
We emphasize that the probabilities for classical field configurations are positive. 
This distinguishes our approach from the Wigner function.

\indent The embedding of a quantum system into a classical probabilistic system
has often been considered as being impossible. 
The pioneering work of Koopman \cite{KOOP} and von Neumann \cite{VNEU} on a Hilbert space formulation of classical mechanics has become close to a quantum formalism for classical statistics by Scho\"nberg's introduction of a complex wave function \cite{MS1,MS2,ALO,RWI,SUD,BAR} whose absolute square yields the probability. 
This has led to a large number of interesting developments by the use of operator methods in classical statistical systems \cite{KAN, MAMA, MAU, GOMA, NIC, NAKL, VOLO, BON, AME, KPM, ACKA, CHRU, KLE, MEZI, DW, SGK, NAME, KKB, HAV, NOJO, MEZI1, MEZI2, MAUROY, VAGI}. 
Nevertheless, several points have led to the assertion that these systems do not account for generic quantum systems. 
First, the operators for classical observables all commute. 
Second, the phases in the Koopman--von Neumann--Scho\"nberg (KNS) wave function play no dynamical role.
Third, general no-go theorems such as Bell's inequalities \cite{BEL1, CHSH} for classical correlation functions seem to forbid the description of genuine quantum systems by classical statistical systems.

\indent A central new point in our approach is the use of statistical observables \cite{CWQFTCF, CWTE, CWQO}. 
Indeed, the conserved charges are statistical observables which do not take a fixed value for a given field configuration. 
They rather characterize properties of the probability distribution for field configurations
with respect to phase rotations for the complex field. 
Statistical observables are well known in classical statistics. 
Prominent examples are temperature or pressure in an equilibrium ensemble. 
These observables can definitely be measured. 
They do not take a fixed value for given microstates, however. 
For example, they are not functions of the positions and momenta of the molecules in a gas. 
Temperature and pressure characterize properties of the probability distribution. 
Other examples are the local density of a fluid or macroscopic correlation functions. 
We will argue that statistical observables play a central role for microphysics, or for the properties of quantum particles.

\indent The presence of statistical observables changes the view of probabilistic classical field theories in important aspects. 
Classical correlation functions are not defined for pairs of statistical observables or statistical and classical observables. 
Simultaneous probabilities for pairs of values for such observables do not exist. 
Sequences of measurements of statistical observables are described by correlation functions which are not based on simultaneous probabilities -- they rather involve conditional probabilities \cite{CWPW}.
Simultaneous probabilities are a central assumption of Bell's inequalities. 
Since this assumption is not realized, one concludes that Bell's inequalities do not need to hold for possible correlations involving statistical observables. 
This lack of simultaneous probabilities is closely related to a second important property of statistical observables: 
The operators representing statistical observables do not commute with the ones for classical observables. 
Also operators for two statistical observables do often not commute.

\indent A second important difference between our approach and the KNS -- wave function is the use of a real classical wave function \cite{CWQP1, CWIT, CWQFC} whose components are simply the square roots of the probabilities. 
Our starting point are real quantities, not complex numbers with phases.
Still, one can introduce a complex structure by mapping the real wave function to a complex wave function, or expressing a real wave function or probability distribution in terms of a complex wave function, as in eq.~\eqref{eq:I2}.
For this type of complex wave function the probability distribution depends on the phases of the complex wave function. 
The expectation values of both classical and statistical observables depend on these phases.
Thus the phases are crucial for the dynamics and observation, in contrast to the KNS--wave function for which the phases are redundant. 
These phases play a crucial role for quantum mechanical interference as for the double slit experiment.

\indent The purpose of this paper is not an abstract discussion of concepts.
We rather present a concrete computation how the quantum particle in a potential emerges from a 
probabilistic classical field theory.
We proceed in a rather detailed manner which highlights the emergence of the quantum concepts
in a classical statistical setting.

\subsection*{Quantum field theory as subsystem of probabilistic classical field theory}

\indent We consider a complex classical scalar field $\sigma(t,\vec x)$ which obeys a 
deterministic field equation with a linear force,
\begin{equation}
\partial_t \sigma(t,\vec x) = \pi(t,\vec x), \quad
\partial_t \pi(t,\vec x) = - B \sigma(t,\vec x)\,.
\label{eq:1}
\end{equation}
Here $B$ is a real symmetric positive operator which we take here independent of $t$.
A typical example is
\begin{equation}
B = -\partial_x^2 + m^2\,,
\label{eq:2}
\end{equation}
where $\partial_x^2$ stands for the Laplace operator, $\partial_x^2 = \partial_k \partial_k$,  $\partial_k = \partial/\partial x_k$,
with summation over double indices always implied. 
The example \eqref{eq:2} corresponds to a free relativistic complex scalar field with mass $m$.
With probabilistic initial conditions we deal with a classical statistical system. 
The probability distribution $w(t;\sigma,\pi)$
indicates at every time $t$ the probability to find the field configuration 
$(\sigma(\vec x),\pi(\vec x))$. 
Its evolution obeys the Liouville equation
\begin{equation}
\partial_t w = -\hat L w, \quad
\hat L = \hat L_K + \hat L_V\,,
\label{eq:3}
\end{equation}
with Liouville operator $\hat L$ given by
\begin{align}
\hat L_K &=
\int_{\vec x}
\left(
\pi(\vec x)\frac{\partial}{\partial \sigma(\vec x)}
+ \pi^*(\vec x)\frac{\partial}{\partial \sigma^*(\vec x)}
\right)\,,
\nonumber\\
\hat L_V &=
-\int_{\vec x}
\left(
B\,\sigma(\vec x)\frac{\partial}{\partial \pi(\vec x)}
+ B\,\sigma^*(\vec x)\frac{\partial}{\partial \pi^*(\vec x)}
\right)\,.
\label{eq:4}
\end{align}
We will equivalently use a description in terms of two real fields $\sigma_1$ and $\sigma_2$,
\begin{equation}
\sigma = \frac{1}{\sqrt2}(\sigma_1+i\sigma_2), \quad
\pi = \frac{1}{\sqrt2}(\pi_1+i\pi_2)\,.
\label{eq:5}
\end{equation}
In this formulation the Liouville operator reads
\begin{equation}
\hat L =
\sum_{j=1,2}\int_{\vec x}
\left(
\pi_j(\vec x)\frac{\partial}{\partial \sigma_j(\vec x)}
- B\sigma_j(\vec x)\frac{\partial}{\partial \pi_j(\vec x)}
\right)\,.
\label{eq:6}
\end{equation}

\indent We work with the real classical wave function \cite{CWQP1, CWIT, CWQFC}, 
which is the square root of the probability distribution,
\begin{equation}
w(t;\sigma,\pi) = q^2(t;\sigma,\pi)\,.
\label{eq:7}
\end{equation}
Its evolution obeys the Liouville equation as well,
\begin{equation}
\partial_t q = -\hat L q\,.
\label{eq:8}
\end{equation}
Eq.~\eqref{eq:8} can be written in the form of a Schr\"odinger equation,
\begin{equation}
i\partial_t q = H_L q, \quad
H_L = -i\hat L, \quad
H_L^\dagger = H_L\,.
\label{eq:9}
\end{equation}
We express $H_L$ in terms of hermitian field operators,
\begin{equation}
H_L =
\sum_j \int_{\vec x}
\left[
\hat\pi_j(\vec x)\hat\gamma_j(\vec x)
+
\bigl(B\hat\sigma_j(\vec x)\bigr)\hat\zeta_j(\vec x)
\right]\,.
\label{eq:11}
\end{equation}
In the basis where $q$ is a function of $\sigma_j(\vec x)$ and $\pi_j(\vec x)$ 
these operators are given by
\begin{align}
\hat\sigma_j(\vec x) &= \sigma_j(\vec x)\,, &
\hat\gamma_j(\vec x) &= -i\frac{\partial}{\partial \sigma_j(\vec x)}\,, \nonumber\\
\hat\pi_j(\vec x) &= \pi_j(\vec x)\,, &
\hat\zeta_j(\vec x) &= i\frac{\partial}{\partial \pi_j(\vec x)}\,.
\label{eq:12}
\end{align}

\indent Defining the linear combinations
\begin{align}
\hat\varphi_j(\vec x) &= \hat\sigma_j(\vec x) + \frac{1}{2}\hat\zeta_j(\vec x), &
\hat\chi_j(\vec x) &= \hat\sigma_j(\vec x) - \frac{1}{2}\hat\zeta_j(\vec x), \nonumber\\
\hat\eta_j(\vec x) &= \hat\pi_j(\vec x) + \frac{1}{2}\hat\gamma_j(\vec x), &
\hat\beta_j(\vec x) &= \hat\pi_j(\vec x) - \frac{1}{2}\hat\gamma_j(\vec x)\,,
\label{eq:13}
\end{align}
the Hamilton operator decomposes into two independent parts,
\begin{align}
H_L &= H^{(q)} - H^{(m)}\,, \nonumber\\
H^{(q)} &=
\frac12 \sum_j \int_{\vec x}
\left(
\hat\eta_j^2(\vec x)
+ \hat\varphi_j(\vec x)\,B\,\hat\varphi_j(\vec x)
\right)\,, \nonumber\\
H^{(m)} &=
\frac12 \sum_j \int_{\vec x}
\left(
\hat\beta_j^2(\vec x)
+ \hat\chi_j(\vec x)\,B\,\hat\chi_j(\vec x)
\right)\,.
\label{eq:14}
\end{align}
The set of operators $(\hat\varphi_j,\hat\eta_j)$ commutes with all operators in the set $(\hat\chi_j,\hat\beta_j)$. 
The non-zero commutation relations in each sector are
\begin{align}
[\hat\varphi_i(\vec x),\hat\eta_j(\vec y)] &=
i\delta_{ij}\delta(\vec x-\vec y), \nonumber\\
[\hat\chi_i(\vec x),\hat\beta_j(\vec y)] &=
-i\delta_{ij}\delta(\vec x-\vec y)\,.
\label{eq:15}
\end{align}

\indent The operator $\dot{\hat A}$ for the time derivative of an observable $A$ is given by the commutator with $H_L$,
\begin{equation}
\dot{\hat A} = i[H_L,\hat A]\,.
\label{eq:16}
\end{equation}
From eq.~\eqref{eq:14} one infers
\begin{align}
\dot{\hat\varphi}_j(\vec x) = \hat\eta_j(\vec x)\,, \quad
\dot{\hat\eta}_j(\vec x) = -B\hat\varphi_j(\vec x)\,.
\label{eq:17}
\end{align}
These are the evolution equations for the field operators of a standard quantum field theory for a complex scalar field. 
The field operators $\hat\varphi(\vec x)$, $\hat\eta(\vec x)$ form a closed subsystem of the more general classical statistical system. 

\indent We may consider $m^2(\vec x)$ depending on $\vec x$ and the ``non-relativistic limit''  
$-\partial_x^2 \ll m^2$. In this limit it is well known that this quantum field theory yields
for the one-particle excitations the Schr\"odinger equation for a quantum particle in a 
potential. For a suitable potential this realizes the setting for the double-slit experiment 
in quantum mechanics. In the following we recapitulate this result in the context of a classical field theory.
An important role is played by the conserved charges. Its integer values provide for the 
discreteness of the outcome of measurements -- the discrete clicks in particle detectors.

\subsection*{Change of field-basis}

\indent An important advantage of the quantum formalism for classical statistics \cite{CWQFC} is the possibility to express the operator 
relations in an arbitrary basis. We may choose a basis of eigenvectors of $\hat\varphi_i$, 
for which the operators take the form
\begin{equation}
\hat\varphi_j(x)=\varphi_j(x), \quad
\hat\eta_j(x)=-i\frac{\partial}{\partial \varphi_j(x)}\,.
\label{eq:18}
\end{equation}
In this basis one has
\begin{align}
H^{(q)}
=
\frac12\sum_j \int_{\vec x,\vec y}
\Bigg\{
&-\frac{\partial}{\partial \varphi_j(\vec x)}
\,\delta(\vec x-\vec y)\,
\frac{\partial}{\partial \varphi_j(\vec y)}
\nn \\
&+
\varphi_j(\vec x) B(\vec x,\vec y)\varphi_j(\vec y)
\Bigg\}\,.
\label{eq:19}
\end{align}
(One could generalize eq.~\eqref{eq:1} and $\hat L_K$ with $\delta(\vec{x}-\vec{y})$ replaced by a
more general symmetric function.) 
The operator $B$ needs not be homogeneous in space. For example, we may take in eq.~\eqref{eq:2}
\begin{equation}
m(\vec x) = m + V(\vec x)\,.
\label{eq:19A}
\end{equation}
This results in
\begin{align}
\label{eq:20}
H^{(q)}
=
&\frac12\sum_j \int_{\vec x}
\Bigg\{
-\frac{\partial^2}{\partial \varphi_j^2(\vec x)}
\\
&+
\varphi_j(\vec x)\bigl((m+V(\vec x))^2-\partial_x^2\bigr)\varphi_j(\vec x)
\Bigg\}\nn\,.
\end{align}

\indent The ``mirror part'' $H^{(m)}$ of the Hamiltonian \eqref{eq:14} obtains from $H^{(q)}$ by a 
transformation related to time reversal,
\begin{align}
\label{eq:21}
\hat T:\quad\,\,
&\hat\sigma\to\hat\sigma\,,\quad
\hat\pi\to-\hat\pi\,,\quad
\hat\gamma\to\hat\gamma\,,\quad
\hat\zeta\to -\hat\zeta\,, \\
&\hat\varphi \leftrightarrow \hat\chi\,, \quad
\hat\eta\leftrightarrow\hat\beta,\quad
H^{(q)}\leftrightarrow H^{(m)}\,, \quad
H_L\to -H_L\,. \nn
\end{align}
The field operators $(\hat\chi,\hat\beta)$ obey the same evolution equation \eqref{eq:17} 
as the field operators $(\hat\varphi,\hat\eta)$.
Observables of the classical field theory beyond the subsystem of the quantum field theory can be 
constructed as functions of $\hat\varphi$, $\hat\eta$, $\hat\chi$ and $\hat\beta$. 

\indent The expectation values of all operators built from $\hat\varphi_j(\vec x)$ and
$\hat\eta_j(\vec x)$ follow precisely the evolution of quantum field theory with Hamiltonian $H^{(q)}$.
This is the central result of this note. 
From there we need to construct the one-particle states and take the non-relativistic limit. 
Its time evolution will obey the Schr\"odinger equation for a quantum particle in an arbitrary potential $V(\vec x)$.

\indent The wave function in the basis \eqref{eq:18} obtains by a functional Fourier transform,
\begin{equation}
\tilde\psi(\sigma,\zeta)
=
\int \hat{\mathcal D}\pi\,
\exp\left\{
i\sum_j \int_{\vec x}\zeta_j(\vec x)\pi_j(\vec x)
\right\}\,
q(\sigma,\pi)\,,
\label{eq:22}
\end{equation}
where
\begin{equation}
\int \hat{\mathcal D}\pi =
\prod_{j} \prod_{\vec x}\left(  \frac{1}{2\pi}\int_{-\infty}^{\infty}
d\pi_j(\vec x)\right)\,.
\label{eq:23}
\end{equation}
In this basis one has  $\hat\zeta_j(\vec x)=\zeta_j(\vec x)$, one finds eq.~\eqref{eq:18}, and
$\hat \chi_j (\vec x) =  \chi_j (\vec x)$, $\hat \beta_j (\vec x) = i \partial/\partial \chi_j (\vec x)$.
For a real wave function $q(\sigma,\pi)$ the complex wave function $\tilde\psi(\sigma,\zeta)$ obeys the constraint 
\begin{equation}
\tilde\psi(\sigma,-\zeta)=\tilde\psi^\ast(\sigma,\zeta)\,,
\quad
\tilde\psi^\ast(\varphi,\chi)=\tilde\psi(\chi,\varphi)\,.
\label{eq:24}
\end{equation}

\indent We will be mainly concerned here with ``pure state wave functions'' for which $\tilde\psi$ factorizes as
\begin{equation}
\tilde\psi(\varphi,\chi)=\psi(\varphi)\psi^*(\chi)\,.
\label{eq:25}
\end{equation}
The constraint \eqref{eq:24} is obeyed for arbitrary complex wave functions $\psi (\varphi)$. 
The evolution generator for $\psi(\varphi)$ is $H^{(q)}$,
\begin{equation}
i\partial_t \psi(\varphi)=H^{(q)}\psi(\varphi)\,.
\label{eq:26}
\end{equation}
In turn, every solution of eq.~\eqref{eq:26} solves, with eq.~\eqref{eq:25}, 
the evolution equation for $\tilde\psi$,
\begin{equation}
i\partial_t \tilde\psi(\sigma,\zeta)
=
\bigl(H^{(q)}-H^{(m)}\bigr)\tilde\psi(\sigma,\zeta)\,,
\label{eq:27}
\end{equation}
which is the functional Fourier transform of eq.~\eqref{eq:9}.
For every solution of eq.~\eqref{eq:26} we can therefore construct $q(\sigma,\pi)$ by inverting 
the Fourier transform \eqref{eq:22}. 
The real wave function $q$ obeys the Liouville equation \eqref{eq:8}.
In turn, the probability distribution $w=q^2$ obeys eq.~\eqref{eq:3}. 
We conclude that every solution of the complex Schr\"odinger equation \eqref{eq:26} 
for $\psi$ constitutes a solution of the Liouville equation for the associated probability distribution $w$. 
We emphasize that the superposition of two solutions of the Schr\"odinger equation \eqref{eq:26} is again a solution of this equation. 
The corresponding probability distribution is, however, not a simple addition of the associated probability distributions. 
One observes the interference characteristic for quantum mechanics. 
A general solution of the Liouville equation \eqref{eq:8} for $q$ can be written as a linear combination of pure state wave functions \eqref{eq:25}. 
This results in a density matrix for the description of observables constructed from the field operators $\hat\varphi$ and $\hat\eta$.

\subsection*{Fluctuating fields}

\indent The field operators $\hat{\varphi}(\vec{x})$ are associated to field observables $\varphi(\vec{x})$ for ``fluctuating fields'' \cite{CWQFTCF}.
They are statistical observables, as seen easily from the representation of the field operators in the $q$-basis
\begin{equation}
\hat{\varphi}_j(\vec{x})
=
\sigma_j(\vec{x})
+
\frac{i}{2}
\frac{\partial}{\partial\pi_j(\vec{x})}\,.
\label{eq:FF1}
\end{equation}
A typical probability distribution for a microphysical description shows a substantial dispersion in the time derivatives $\pi(\vec{x})$.
As a consequence, precise measurements of local field variables $\sigma(\vec{x})$ are practically impossible -- a precise field value changes in a time which is much shorter than any characteristic experimental time.
The fluctuating field observables $\varphi(\vec{x})$ are more robust. 
They incorporate the genuine uncertainty due to the dispersion in $\pi(\vec{x})$. 
They do not assume arbitrarily sharp values, as can be seen from
the relation
\begin{equation}
\langle \varphi^2(\vec{x})\rangle
-
\langle \varphi(\vec{x})\rangle^2
=
\langle \sigma^2(\vec{x})\rangle
-
\langle \sigma(\vec{x})\rangle^2
+
\frac14
\langle \zeta^2(\vec{x})\rangle\,,
\label{eq:FF2}
\end{equation}
while
\begin{equation}
\langle \varphi(\vec{x})\rangle
=
\langle \sigma(\vec{x})\rangle\,.
\label{eq:FF3}
\end{equation}
The statistical observable $\zeta^2(\vec{x})$ measures the roughness of the probability distribution in the time derivatives of the classical fields
\begin{equation}
\langle \zeta^2(\vec{x})\rangle
=
\int_{\sigma,\pi}
\left(
\frac{\partial q}
{\partial\pi(\vec{x})}
\right)^2\,.
\label{eq:FF4}
\end{equation}
With $\eta(\vec{x})$ the statistical observable for the time-derivative of the fluctuating field according to eq.~\eqref{eq:17}, the commutator relation \eqref{eq:15} enforces the characteristic uncertainty relation for $\varphi(\vec{x})$ and $\eta(\vec{x})$.
One sees the emergence of typical quantum structures in our classical field theory.
Changing units of $H_L$ from inverse time to some other energy unit Planck's constant $\hbar$ appears in eqs.~\eqref{eq:16}, \eqref{eq:17}.

\subsection*{Conserved charge}

\indent In the following we focus on solutions of the Schr\"odinger equation \eqref{eq:26} with Hamiltonian \eqref{eq:19}. 
We next discuss the conserved charge. 
The Hamiltonian $H^{(q)}$ is invariant under rotations among the two components $\varphi_1$ and $\varphi_2$. 
An infinitesimal rotation, 
\begin{equation}
\delta\varphi_1(\vec x)=-\delta\beta\,\varphi_2(\vec x)\,,\quad
\delta\varphi_2(\vec x)=\delta\beta\,\varphi_1(\vec x)\,,
\label{eq:28}
\end{equation}
results in the change of the wave function 
\begin{align}
\delta\psi
&=
\delta\beta \int_{\vec x}
\left(
\varphi_1(\vec x)\frac{\partial}{\partial \varphi_2(\vec x)}
-
\varphi_2(\vec x)\frac{\partial}{\partial \varphi_1(\vec x)}
\right)\psi
\nn\\
&=
-i\delta\beta\,\hat Q\,\psi\,.
\label{eq:29}
\end{align}
The charge operator is the generator of these "flavor rotations",
\begin{equation}
\hat Q
=
\int_{\vec x}
\left(
\hat\varphi_2(\vec x)\hat\eta_1(\vec x)
-
\hat\varphi_1(\vec x)\hat\eta_2(\vec x)
\right)\,.
\label{eq:30}
\end{equation}
It commutes with $H^{(q)}$,
\begin{equation}
[\hat Q,H^{(q)}]=0\,,
\label{eq:31}
\end{equation}
such that the charge $Q$ is a conserved quantity.
The charge operator obeys the commutation relations
\begin{align}
[\hat Q,\hat\varphi_1(\vec x)] &= -i\hat\varphi_2(\vec x)\,, &
[\hat Q,\hat\varphi_2(\vec x)] &= i\hat\varphi_1(\vec x)\,, \nonumber\\
[\hat Q,\hat\eta_1(\vec x)] &= -i\hat\eta_2(\vec x)\,, &
[\hat Q,\hat\eta_2(\vec x)] &= i\hat\eta_1(\vec x)\,.
\label{eq:31A}
\end{align}

\indent The charge is a statistical observable, since the associated operator $\hat Q$ involves derivatives 
with respect to $\sigma$ and $\pi$. It does not take fixed values for the ``microstates'' 
which correspond to field configurations $(\sigma(\vec x),\pi(\vec x))$. 
It rather measures properties of the probabilistic information encoded in the wave function
which are related to flavor rotations between the two real components.

\indent The charge operator is hermitian, $\hat Q^\dagger=\hat Q$. 
Its eigenvalues are integers. 
This is seen most clearly if we employ a complex field,
\begin{equation}
\varphi = \frac{1}{\sqrt2}(\varphi_1+i\varphi_2)\,.
\label{eq:32}
\end{equation}
A flavor rotation \eqref{eq:28} corresponds to a global phase change of the complex field,
\begin{equation}
\varphi^\prime = e^{i\beta}\varphi\,.
\label{eq:33}
\end{equation}
According to eq.~\eqref{eq:29} the change of the wave function $\psi$ under a rotation of $\varphi$ obeys
\begin{equation}
\frac{\partial\psi}{\partial\beta}=-i\hat Q\psi\,.
\label{eq:34}
\end{equation}
For an eigenstate of $\hat Q$ one has
\begin{equation}
\hat Q\psi_Q=Q\psi_Q\,,
\label{eq:44A}
\end{equation}
and the rotated eigenstate obeys
\begin{equation}
\psi_Q(\beta)=\exp(-i\beta Q)\psi_Q(0)\,.
\label{eq:35}
\end{equation}
Under a rotation with $\beta=2\pi$ the field configuration does not change and therefore $\psi(\beta=2\pi)=\psi(0)$. 
This implies that the eigenvalues $Q$ are integers.

\indent Expressed in terms of the complex field $\varphi$ one has
\begin{equation}
\hat Q
=
\int_{\vec x}
\left(
\varphi^*(\vec x)\frac{\partial}{\partial\varphi^*(\vec x)}
-
\varphi(\vec x)\frac{\partial}{\partial\varphi(\vec x)}
\right)\,.
\label{eq:36}
\end{equation}
Under complex conjugation the charge operator changes sign
\begin{equation}
\hat Q^*=-\hat Q\,.
\label{eq:36A}
\end{equation}
In consequence, the wave function $\psi_Q^*$ is an eigenstate with eigenvalue $-Q$,
\begin{equation}
\hat Q\psi_Q^*=-Q\psi_Q^*\,.
\label{eq:36B}
\end{equation}

\indent For the complex field operators,
\begin{align}
&\hat\varphi(\vec x)=\frac{1}{\sqrt2}\bigl(\hat\varphi_1(\vec x)+i\hat\varphi_2(\vec x)\bigr)\,,
\nn \\
&\hat\eta(\vec x)=\frac{1}{\sqrt2}\bigl(\hat\eta_1(\vec x)+i\hat\eta_2(\vec x)\bigr)\,,
\label{eq:36C}
\end{align}
one infers the commutation relations
\begin{align}
[\hat Q,\hat\varphi(\vec x)] &= -\hat\varphi(\vec x)\,, &
[\hat Q,\hat\varphi^\dagger(\vec x)] &= \hat\varphi^\dagger(\vec x)\,, \nonumber\\
[\hat Q,\hat\eta(\vec x)] &= -\hat\eta(\vec x)\,, &
[\hat Q,\hat\eta^\dagger(\vec x)] &= \hat\eta^\dagger(\vec x)\,.
\label{eq:37}
\end{align}
If $\hat\varphi^\dagger\psi_Q$ does not vanish it is an eigenstate of $\hat Q$ with eigenvalue $Q+1$,
\begin{equation}
\hat Q\,\hat\varphi^\dagger\psi_Q
=
(Q+1)\hat\varphi^\dagger\psi_Q\,.
\label{eq:37A}
\end{equation}
This holds similarly for $\hat\eta^\dagger\psi_Q$, while $\hat\varphi\psi_Q$ and $\hat\eta\psi_Q$ 
are eigenstates with eigenvalues $Q-1$. This shows that $\hat\varphi$, $\hat\eta$, $\hat\varphi^\dagger$
and $\hat\eta^\dagger$ are related to "annihilation and creation operators" which lower or increase the charge. 
We will make this relation more precise below.

\subsection*{Vacuum}

\indent The quantum vacuum is given by a homogeneous wave function $|0\rangle$ which is an eigenstate 
of $\hat Q$ with eigenvalue zero. One-particle excitations have $Q=1$, and higher values of $Q$ 
correspond to multi-particle excitations. Antiparticles are related to negative values of $Q$.
The vacuum should have the property that all particle or antiparticle excitations have 
a bounded quantum energy, $\langle H^{(q)}\rangle > E_0$. With $H^{(q)}|0\rangle = E_0 |0\rangle $
this guarantees stability of the vacuum. 
From there we will construct the one-particle states by applying a creation operator on the vacuum.

\indent We first perform the familiar construction for constant $m^2$, corresponding to the quantum 
field theory for a relativistic free complex scalar field. More general $B$ will be rather similar. 
For constant $m^2$ we switch from fields depending on $\vec x$ to fields depending on the momentum 
$\vec p$ by a Fourier transform, $p=-i\partial_x$,
\begin{equation}
\label{eq:38}    
B=\vec p^{\,2}+m^2\,.
\end{equation}
In this basis one has
\begin{equation}
H^{(q)}
=
\int_{\vec p}
\left(
\hat\eta^\dagger(\vec p)\hat\eta(\vec p)
+
B(\vec p)\hat\varphi^\dagger(\vec p)\hat\varphi(\vec p)
\right)\,,
\label{eq:39}
\end{equation}
and
\begin{equation}
\hat Q
=
-i\int_{\vec p}
\left(
\hat\varphi(\vec p)\hat\eta^\dagger(\vec p)
-
\hat\varphi^\dagger(\vec p)\hat\eta(\vec p)
\right)\,.
\label{eq:40}
\end{equation}
(Factors of $2\pi$ characteristic for Fourier transforms are incorporated in $\int_{\vec p}$ and $\delta(\vec p-\vec q)$.) 
The commutation relations \eqref{eq:37} take over with $\vec x$ replaced by $\vec p$, where
\begin{equation}
[\hat\varphi(\vec p),\hat\eta^\dagger(\vec q)]
=
i\delta(\vec p-\vec q)\,,
\quad
[\hat\varphi(\vec p),\hat\eta(\vec q)]=0\,.
\label{eq:41}
\end{equation}

\indent We introduce annihilation and creation operators,
\begin{align}
a(\vec p) &=
\frac{1}{\sqrt2}
\left(
\alpha(\vec p)\hat\varphi(\vec p)
+
\frac{i}{\alpha^*(\vec p)}\hat\eta(\vec p)
\right)\,, \nonumber\\
a^\dagger(\vec p) &=
\frac{1}{\sqrt2}
\left(
\alpha^*(\vec p)\hat\varphi^\dagger(\vec p)
-
\frac{i}{\alpha(\vec p)}\hat\eta^\dagger(\vec p)
\right)\,.
\label{eq:42}
\end{align}
They obey the commutation relation
\begin{equation}
[a(\vec p),a^\dagger(\vec q)]
=
\delta(\vec p-\vec q)\,, \quad
[a(\vec p),a(\vec q)]=0\,,
\label{eq:43}
\end{equation}
and the anticommutation relation
\begin{align}
\{a(\vec p),a^\dagger(\vec p)\}
=
&|\alpha(\vec p)|^2\hat\varphi^\dagger(\vec p)\hat\varphi(\vec p)
\nn\\
&+
\frac{1}{|\alpha(\vec p)|^2}\hat\eta^\dagger(\vec p)\hat\eta(\vec p)
+
\hat Q(\vec p)\,,
\label{eq:44}
\end{align}
with
\begin{equation}
\hat Q(\vec p)
=
-i\left(
\hat\varphi(\vec p)\hat\eta^\dagger(\vec p)
-
\hat\varphi^\dagger(\vec p)\hat\eta(\vec p)
\right)\,,
\quad
\int_{\vec p}\hat Q(\vec p)=\hat Q\,.
\label{eq:45}
\end{equation}
From the commutation relations,
\begin{equation}
[\hat Q,a(\vec p)] = -a(\vec p)\,,
\quad
[\hat Q,a^\dagger(\vec p)] = a^\dagger(\vec p)\,,
\label{eq:46}
\end{equation}
one infers that the annihilation or creation operators lower or increase the charge, respectively,
\begin{align}
\hat Q\bigl(a(\vec p)\psi_Q\bigr)
&=
(Q-1)\bigl(a(\vec p)\psi_Q\bigr)\,,
\nonumber\\
\hat Q\bigl(a^\dagger(\vec p)\psi_Q\bigr)
&=
(Q+1)\bigl(a^\dagger(\vec p)\psi_Q\bigr)\,.
\label{eq:47}
\end{align}
(This assumes that $a(\vec p)\psi_Q$ or $a^\dagger(\vec p)\psi_Q$ do not vanish.)

\indent A second set of annihilation and creation operators is defined by
\begin{align}
c(\vec p) &=
\frac{1}{\sqrt2}
\left(
\gamma(\vec p)\hat\varphi(\vec p)
-
\frac{i}{\gamma^*(\vec p)}\hat\eta(\vec p)
\right)\,, \nonumber\\
c^\dagger(\vec p) &=
\frac{1}{\sqrt2}
\left(
\gamma^*(\vec p)\hat\varphi^\dagger(\vec p)
+
\frac{i}{\gamma(\vec p)}\hat\eta^\dagger(\vec p)
\right)\,.
\label{eq:48}
\end{align}
It obeys
\begin{align}
[c(\vec p),c^\dagger(\vec q)] &= -\delta(\vec p-\vec q)\,, \nonumber\\
\{c(\vec p),c^\dagger(\vec p)\}
&=
|\gamma(\vec p)|^2\hat\varphi^\dagger(\vec p)\hat\varphi(\vec p)
\nn\\
&+
\frac{1}{|\gamma(\vec p)|^2}\hat\eta^\dagger(\vec p)\hat\eta(\vec p)
-
\hat Q(\vec p)\,.
\label{eq:49}
\end{align}
For the choice $|\gamma(\vec p)|^2=|\alpha(\vec p)|^2$
the operators $c$ and $c^\dagger$ commute with $a$ and $a^\dagger$ and one finds 
\begin{equation}
\hat Q(\vec p)
=
a^\dagger(\vec p)a(\vec p)
-
c(\vec p)c^\dagger(\vec p)\,.
\label{eq:50}
\end{equation}
We further choose
\begin{equation}
|\alpha(\vec p)|^2
=
|\gamma(\vec p)|^2
=
\sqrt{B(\vec p)}
=
\omega(\vec p)\,.
\label{eq:51}
\end{equation}
This expresses the Hamiltonian as
\begin{equation}
H^{(q)}
=
\int_{\vec p}
\omega(\vec p)
\left(
a^\dagger(\vec p)a(\vec p)
+
c^\dagger(\vec p)c(\vec p)
\right)\,.
\label{eq:52}
\end{equation}

\indent We interpret $a^\dagger(\vec p)$ as the creation operator for a particle with charge $Q=1$, 
and $b^\dagger (\vec p) = c (\vec p)$ as the creation operator for an antiparticle with charge $Q=-1$,
\begin{align}
&b(\vec p)=c^\dagger(\vec p)\,,\quad
b^\dagger(\vec p)=c(\vec p)\,,\quad \nn\\
&\hat Q(\vec p)=a^\dagger(\vec p)a(\vec p)-b^\dagger(\vec p)b(\vec p)\,.
\label{eq:67A}
\end{align}

\indent The Hamiltonian of the quantum field theory takes the form
\begin{equation}
H^{(q)} = \int_{\vec p} \omega (\vec p) \left( a^\dagger(\vec{p})a(\vec{p}) + b^\dagger(\vec{p})b(\vec{p})  \right) + E_0 \,.
\label{eq:53}
\end{equation}
The ``vacuum energy'' $E_0$ arises from reordering the operators $c^\dagger$ and $c$ in eq.~\eqref{eq:52}
\begin{equation}
E_0
=
\int_{\vec p}\omega(\vec p)[b(\vec p),b^\dagger(\vec p)]
=
\Omega \int_{\vec p}\omega(\vec p)\,,
\label{eq:54}
\end{equation}
where $\Omega=\int_{\vec x}$ is the volume of space. With
\begin{equation}
[H^{(q)},a^\dagger(\vec p)]
=
\omega(\vec p)a^\dagger(\vec p)\,,
\quad
[H^{(q)},b^\dagger(\vec p)]
=
\omega(\vec p)b^\dagger(\vec p)\,,
\label{eq:55}
\end{equation}
the creation of a particle or an antiparticle increases the quantum energy by $\omega(\vec p)$.

\indent A vacuum state with zero charge is realized by the condition
\begin{equation}
a(\vec p)\,|0\rangle=0\,,
\quad
b(\vec p)\,|0\rangle=0\,.
\label{eq:56}
\end{equation}
We can interpret
$\hat N_p=\int_{\vec p}a^\dagger(\vec p)a(\vec p)$,
$\hat N_a=\int_{\vec p}b^\dagger(\vec p)b(\vec p)$,
as the number operators for particles and antiparticles, respectively.
 The vacuum has zero particle and antiparticle number,
\begin{equation}
\hat N_p\,|0\rangle=0\,,\quad
\hat N_a\,|0\rangle=0\,,\quad
\hat Q=\hat N_p-\hat N_a\,.
\label{eq:57}
\end{equation}

One can construct a basis for the wave function $\psi$ by applying products of creation operators 
$a^\dagger(\vec p)$ or $b^\dagger(\vec p)$ on the vacuum state.
These are eigenstates of $H^{(q)}$. 
The eigenvalues of $H^{(q)}-E_0$ are sums of frequencies $\omega(\vec p)$, with a term $\omega(\vec p)$
for each factor $a^\dagger(\vec p)$ or $b^\dagger(\vec p)$. 
They are all positive. 
One concludes that with the vacuum condition \eqref{eq:56} the vacuum state $|0\rangle$ is the one with 
lowest quantum energy. For all excitations of the vacuum the expectation value $\langle H^{(q)}\rangle$ 
is larger than the ground state energy $E_0$. 
The ground state energy is not zero -- for discrete momenta eq.~\eqref{eq:54} becomes 
$E_0 = \sum_p \sqrt{\vec p^{\,2}+m^2}$.
Thus the vacuum wave function $|0\rangle$ oscillates as $\exp(-iE_0t)$. 
An overall oscillation with an overall phase does 
not matter for quantum mechanics. In our setting the wave function
$\tilde\psi$ in eq.~\eqref{eq:25} is static if $\psi(\varphi)=|0\rangle$,
$\psi^\ast(\chi)=|0\rangle^\ast$, since the phases of the two factors cancel.
In consequence, the vacuum corresponds to static $q$, and therefore to a static probability distribution.

\subsection*{One-particle state}

\indent A one-particle state with $Q=1$ is given by the wave function
\begin{equation}
\psi^{(1)}
=
\int_{\vec p}\psi_S(\vec p)\,
a^\dagger(\vec p)|0\rangle\,,
\quad
\hat Q\psi^{(1)}=\psi^{(1)}\,.
\label{eq:58}
\end{equation}
For the quantum energy one finds
\begin{align}
\bigl(H^{(q)}-E_0\bigr)\psi^{(1)}
&=
\int_{\vec p}\omega(\vec p)\psi_S(\vec p)\,
a^\dagger(\vec p)\,|0\rangle
\nonumber\\
&=
\int_{\vec p,\vec q}
H_S(\vec p,\vec q)\psi_S(\vec q)\,
a^\dagger(\vec p)\,|0\rangle\,.
\label{eq:59}
\end{align}
One extracts the reduced Hamilton operator for the sector of a single
particle with charge $Q=1$,
\begin{align}
H_S(\vec p,\vec q)
&=
\omega(\vec p)\delta(\vec p-\vec q)
=
\sqrt{B(\vec p)}\,\delta(\vec p-\vec q)
\nn \\
&=
\sqrt{\vec p^{\,2}+m^2}\,\delta(\vec p-\vec q)\,.
\label{eq:60}
\end{align}
Transforming back to position space this yields the one-particle Hamiltonian for a free particle,
\begin{equation}
H_S
=
\sqrt{m^2-\partial_x^2}\,\delta(\vec x-\vec y)\,.
\label{eq:61}
\end{equation}
The non-relativistic limit expands for $-\partial_x^2\ll m^2$, omitting
the unit matrix $\delta(\vec x-\vec y)$,
\begin{equation}
H_S
=
m-\frac{1}{2m}\partial_x^2\,.
\label{eq:62}
\end{equation}
The constant $m$ yields again only an additive constant. As long as one
focuses on the one-particle state it can be absorbed by a shift $E_0\to E_0+m$.

\subsection*{Quantum particle in a potential}

\indent This line of argument holds for general $B$, as we will show below. 
Replacing $m$ by $m+V(\vec x)$ one obtains
\begin{equation}
H_S
=
\sqrt{\bigl(m+V(\vec x)\bigr)^2-\partial_x^2}
\approx
m+V(\vec x)-\frac{1}{2m}\partial_x^2\,.
\label{eq:63}
\end{equation}
After subtraction of $m$ this is the Hamiltonian for a quantum
particle in a potential $V(\vec x)$. This holds for an arbitrary
potential, in particular for one which realizes the double-slit experiment.
For a given solution of the one-particle Schrödinger equation
\begin{equation}
i\partial_t\psi_S
=
H_S\psi_S\,,
\label{eq:64}
\end{equation}
one can construct $\psi^{(1)}$ similar to eq.~\eqref{eq:58} in an
arbitrary basis, in particular in position space,
\begin{equation}
\psi^{(1)}
=
\int_{\vec x}\psi_S(\vec x)\,
a^\dagger(\vec x)\,|0\rangle\,.
\label{eq:65}
\end{equation}
If the vacuum wave function $|0\rangle$ is normalized, the one-particle
wave function $\psi^{(1)}$ is normalized according to
\begin{equation}
\int \hat{\mathcal D}\varphi\,\psi^{(1)*}\psi^{(1)}=1\,,
\end{equation}
provided that the Schr\"odinger wave function $\psi_S$ is normalized,
\begin{equation}
\int_{\vec x}\psi_S^*(\vec x)\psi_S(\vec x)=1\,.
\end{equation}
Subsequently, one computes from $\psi^{(1)}$ the corresponding wave functions $\tilde \psi$, $q$ and $w$.
We will do this below explicitly.
This demonstrates for arbitrary $V(\vec x)$ the existence of a
probability distribution $w$ for classical fields which realizes the
quantum particle in a potential $V(\vec x)$.

\indent It remains to be shown that the result for the one-particle Hamiltonian,
\begin{equation}
H_S=\sqrt{B}\,,
\label{eq:66}
\end{equation}
holds for an arbitrary positive hermitian and real $B$ in eq.~\eqref{eq:1}. 
For this purpose we observe that for constant $m$ in eq.~\eqref{eq:2} the momentum $\vec p$ labels the
eigenvalues of $B$. All steps of the construction of the vacuum and
one-particle state continue to hold if we replace $\vec p$ by a label
$n$ for the eigenvalues of $B$.
Instead of momentum space we take a basis where $B$ is diagonal
\begin{equation}
B_{mn}=\omega_n^2\delta_{nm}\,.
\label{eq:67}
\end{equation}
Here $n$ may take discrete or continuous values, according to the properties of the spectrum of $B^2$. 
For a positive operator $B$ all eigenvalues $\omega^2_n$ are positive and we choose positive values for $\omega_n$.
For simplicity we employ a discrete notation. 
For the complex field operator \eqref{eq:36C} we define
\begin{align}
\label{eq:68}
&\hat\varphi(\vec x)=\sum_n U_n(\vec x)\hat\varphi_n\,, \quad
\hat\varphi_n=\int_{\vec x}U_n^*(\vec x)\hat\varphi(\vec x)\,,
\\
&\int_{\vec x}U_n^*(\vec x)U_m(\vec x)=\delta_{nm}\,, \quad
\sum_n U_n(\vec x)U_n^*(\vec y)=\delta(\vec x-\vec y)\,, \nn
\end{align}
and similarly for $\hat\varphi_n^\dagger$, $\hat\eta_n$, $\hat\eta_n^\dagger$.
With
\begin{equation}
B_{mn}
=
\int_{\vec x,\vec y}
U_m^*(\vec y)\,B(\vec y,\vec x)\,U_n(\vec x)\,,
\label{eq:69}
\end{equation}
eqs.~\eqref{eq:39}, \eqref{eq:40}, \eqref{eq:41} continue to hold with
$\vec p,\vec q$ replaced by $n$, $m$, $\int_{\vec p}\to\sum_n$, $\delta(\vec p-\vec q)\to\delta_{nm}$ and $B(\vec p)\to\omega_n^2$. 

\indent The annihilation and creation operators $a_n$, $a_n^\dagger$ are constructed as in eq.~\eqref{eq:42}, e.g.
\begin{align}
&a_n=
\frac{1}{\sqrt2}
\left(
\sqrt{\omega_n}\,\hat\varphi_n
+
\frac{i}{\sqrt{\omega_n}}\,\hat\eta_n
\right)\,,
\nn \\
&b_n=
\frac{1}{\sqrt2}
\left(
\sqrt{\omega_n}\,\hat\varphi_n^\dagger
+
\frac{i}{\sqrt{\omega_n}}\,\hat\eta_n^\dagger
\right)\,.
\label{eq:70}
\end{align}
The commutation relations eqs.~\eqref{eq:43}--\eqref{eq:46}, \eqref{eq:49}
for these operators take over with the corresponding replacements.
This includes the ones with the charge operator
\begin{equation}
\hat Q_n
=
-i\bigl(
\hat\varphi_n\hat\eta_n^\dagger
-
\hat\varphi_n^\dagger\hat\eta_n
\bigr)
=
a_n^\dagger a_n-b_n^\dagger b_n\,,
\quad
\hat Q=\sum_n\hat Q_n\,.
\label{eq:71}
\end{equation}
One arrives at
\begin{align}
\label{eq:72}
H^{(q)}
&= \sum_n \left(\omega^2_n \hat{\varphi}^\dagger_n \hat{\varphi}_n  + \hat{\eta}^\dagger_n \hat{\eta}_n \right)
\\
&=
\sum_n
\omega_n
\left(
a_n^\dagger a_n+b_n^\dagger b_n
\right)
+E_0\,,
\quad
E_0=\sum_n\omega_n\,. \nn
\end{align}

\indent With the vacuum condition $a_n\,|0\rangle=b_n\,|0\rangle=0$ the one-particle state is defined by
\begin{equation}
\psi^{(1)}(t)
=
\sum_n\psi_{S n}(t)a_n^\dagger\,|0\rangle
=
\int_{\vec x}\psi_S(t,\vec x)a^\dagger(\vec x)\,|0\rangle\,,
\label{eq:73}
\end{equation}
with
\begin{equation}
\psi_S(t,\vec x)=\sum_n U_n(\vec x)\psi_{S n}(t)\,,
\quad
a^\dagger(\vec x)=\sum_n U_n^*(\vec x)a_n^\dagger\,.
\label{eq:74}
\end{equation}
It obeys the one-particle Schr\"odinger equation,
\begin{align}
i\partial_t\psi^{(1)}
&=
\bigl(H^{(q)}-E_0\bigr)\psi^{(1)} = \sum_n \omega_n \psi_{Sn} a^\dagger_n\, |0\rangle
\nonumber\\
&=
\int_{\vec x,\vec y}
\sum_n
\omega_n U_n^*(\vec y) U_n(\vec x)\,
\psi_S(\vec y)a^\dagger(\vec x)\,|0\rangle\,.
\label{eq:75}
\end{align}
This amounts to
\begin{equation}
i\partial_t\psi_S(\vec x)
=
\int_{\vec y}H_S(\vec x,\vec y)\psi_S(\vec y)\,,
\label{eq:76}
\end{equation}
with
\begin{equation}
H_S(\vec x,\vec y)
=
\sum_{n,m}U_m(\vec x)\omega_n\delta_{mn}U_n^*(\vec y)
=
\sqrt B\,(\vec x,\vec y)\,.
\label{eq:77}
\end{equation}
For the last identity we employ eqs.~\eqref{eq:68}, \eqref{eq:69} for
the operator $\sqrt B$, $(\sqrt B)_{mn}=\omega_n\delta_{mn}$. 
This completes the proof of eq.~\eqref{eq:66} for an arbitrary form of $B$.

\indent We conclude that the quantum particle in an arbitrary potential can be understood as a particular one-particle probability distribution for a classical field theory. 
The Schr\"odinger equation corresponds to the non-relativistic limit of the classical field equation for a complex scalar field with a space-dependent mass term.
The general derivation of the one-particle Schr\"odinger equation \eqref{eq:76}, \eqref{eq:77} from the classical field equations \eqref{eq:1}, \eqref{eq:2} does not need the ability to actually perform a diagonalization of $B$.
It is sufficient to know that $B$ is a hermitian positive operator which can be diagonalized. 

\indent We have seen that the one-particle Hamiltonian $H_S=B^{1/2}$ plays a key role for the dynamics. 
This is perhaps not too surprising once one observes that for the second-order field equation for
a complex field $\sigma$,
\begin{equation}
\partial_t^2\sigma=-B\sigma\,,
\label{eq:77A}
\end{equation}
a family of solutions given by the root of this equation,
\begin{equation}
i\partial_t\sigma=\sqrt B\,\sigma=H_S\sigma\,.
\label{eq:77B}
\end{equation}
This family corresponds to the particular initial condition
\begin{equation}
\pi(\vec x)=-iH_S\sigma(\vec x)\,.
\label{eq:77C}
\end{equation}

\indent A second key ingredient is the use of appropriate operators which
represent statistical observables. The relations between the complex
field operators and the annihilation and creation operators for particles
and antiparticles read
\begin{align}
\hat\varphi(\vec x)
&=
\frac{1}{\sqrt2}
H_S^{-1/2}
\left(a(\vec x)+b^\dagger(\vec x)\right)\,,
\nonumber\\
\hat\eta(\vec x)
&=
-\frac{i}{\sqrt2}
H_S^{1/2}
\left(a(\vec x)-b^\dagger(\vec x)\right)\,.
\label{eq:77D}
\end{align}
These relations can be employed in an arbitrary field-basis.

\subsection*{Classical probabilities for vacuum and one-particle states}

\indent So far we have constructed the vacuum and one-particle system for the subsystem of the quantum
field theory, assuming the pure state wave function \eqref{eq:25}.
We next translate this to the corresponding classical wave function $q$ and probability distribution $w$.
This will yield the classical probability distributions described in the introductory part.

\indent The vacuum condition,
\begin{align}
&a_n\,|0\rangle=b_n\,|0\rangle=|0\rangle\,, \nn \\
&\hat\eta_n\,|0\rangle=i\omega_n\hat\varphi_n\,|0\rangle\,,\quad
\hat\eta_n^\dagger\,|0\rangle=i\omega_n\hat\varphi_n^\dagger\,|0\rangle\,,
\label{eq:78}
\end{align}
can be evaluated in the basis \eqref{eq:18}, with complex operators \eqref{eq:36C} and complex fields 
$\varphi$
\begin{align}
&\hat\varphi_n=\varphi_n\,,\quad
\hat\varphi_n^\dagger=\varphi_n^*\,, \nn \\
&\hat\eta_n = \int_x U_n (\vec x) \hat \eta (\vec x) = -i \int_x U_n (\vec x) \frac{\partial}{\partial \varphi^\ast (\vec x)}
=-i\frac{\partial}{\partial\varphi_n^*}\,,
\nn \\
&\hat\eta_n^\dagger=-i\frac{\partial}{\partial\varphi_n}\,.
\label{eq:79}
\end{align}
The solution of eq.~\eqref{eq:78} is given by the Gaussian
\begin{align}
|0\rangle
&=
\exp\left(-\sum_n\omega_n\varphi_n^*\varphi_n\right)
\nonumber\\
&=
\exp\left(
-\int_{\vec x,\vec y}
\varphi^*(\vec y)H_S(\vec y,\vec x)\varphi(\vec x)
\right)\,,
\label{eq:80}
\end{align}
where we have omitted a multplicative normalization constant.
For the example \eqref{eq:20} this amounts to
\begin{equation}
|0\rangle
=
\exp\left(
-\int_{\vec x}
\left\{
\varphi^*(\vec x)
\sqrt{\bigl(m+V(\vec x)\bigr)^2-\partial_x^2}\,\,
\varphi(\vec x)
\right\}
\right)\,.
\label{eq:81}
\end{equation}

\indent From eq.~\eqref{eq:80} construct the classical wave function $q^{(0)}$ for the vacuum
state. We first compute $\tilde\psi^{(0)}(\varphi,\chi)$ in eq.~\eqref{eq:25},
by multiplying eq.~\eqref{eq:80} with a similar piece for which $\varphi$ is replaced by $\chi$. 
In terms of the real fields $\sigma_j$ and $\zeta_j$ this yields
\begin{equation}
\tilde\psi^{(0)}(\sigma,\zeta)
=
\exp\left(
-\sum_j\int_{\vec x}
\left\{
\sigma_j H_S\sigma_j
+\frac14\zeta_jH_S\zeta_j
\right\}
\right)\,.
\label{eq:82}
\end{equation}
Performing the inverse Fourier transform \eqref{eq:22} one arrives at the
real vacuum wave function (up to normalization)
\begin{align}
q^{(0)}
=
\mathcal N_0
\exp\Bigg(
-\sum_j\int_{\vec x}
&\Big\{
\sigma_j(\vec x)H_S\sigma_j(\vec x)
\nn \\
&+
\pi_j(\vec x)
H_S^{-1}
\pi_j(\vec x)
\Big\}
\Bigg)\,.
\label{eq:83}
\end{align}
with $\mathcal N_0$ a normalization factor.
One verifies that $q^{(0)}$ obeys the Liouville equation for $\partial_t q_0=0$. The probability distribution for the vacuum is static.
In terms of the complex fields this yields eq.~\eqref{eq:I4}.

\indent We can write the one-particle wave function in an arbitrary field-basis by using the complex field operators,
\begin{equation}
\tilde\psi^{(1)}
=
\int_{\vec x,\vec y}
\psi_S(\vec x)\psi_S^*(\vec y)
\hat C(\vec x,\vec y)\tilde\psi^{(0)}\,,
\label{eq:B10}
\end{equation}
where
\begin{align}
\hat C(\vec x,\vec y)
&=
a^\dagger(\vec x)\tilde a(\vec y)
\nonumber\\
&=
\frac12
\left(
H_S^{1/2}(\vec x)\hat\varphi^\dagger(\vec x)
-iH_S^{-1/2}(\vec x)\hat\eta^\dagger(\vec x)
\right)
\nn \\
&\hspace{0.4cm}\times\left(
H_S^{1/2}(\vec y)\hat\chi(\vec y)
+iH_S^{-1/2}(\vec y)\hat\beta(\vec y)
\right)\,.
\label{eq:B11}
\end{align}
With
\begin{equation}
\check\pi(\vec x)=H_S^{-1}(\vec x)\hat\pi(\vec x)\,,
\quad
\check\gamma(\vec x)=H_S^{-1}(\vec x)\hat\gamma(\vec x)\,,
\label{eq:B12}
\end{equation}
one has
\begin{align}
\hat C(\vec x,\vec y)
&=
H_S^{1/2}(\vec x)
H_S^{1/2}(\vec y)
\tilde C(\vec x,\vec y),
\nonumber\\
\tilde C(\vec x,\vec y)
&=
\frac12
\left(
\hat\sigma^\dagger+\frac12\hat\zeta^\dagger
-i\check\pi^\dagger-\frac i2\check\gamma^\dagger
\right)(\vec x)
\nn \\
&\hspace{4mm}\times\left(
\hat\sigma-\frac12\hat\zeta
+i\check\pi-\frac i2\check\gamma
\right)(\vec y)\,.
\label{eq:B13}
\end{align}

\indent Evaluating this expression in the $(\sigma,\pi)$-basis we employ
\begin{equation}
q^{(0)}
=
\mathcal N_0
\exp\left\{
-2\int_{\vec x}
\left(
\sigma^*(\vec x)H_S\sigma(\vec x)
+
\pi^*(\vec x)H_S^{-1}\pi(\vec x)
\right)
\right\}\,
\label{eq:B14}
\end{equation}
and
\begin{align}
&\hat\zeta(\vec x)=i\frac{\partial}{\partial\pi^*(\vec x)}\,,
\quad
\hat\zeta^\dagger(\vec x)=i\frac{\partial}{\partial\pi(\vec x)}\,,
\nn \\
&\hat\gamma(\vec x)=-i\frac{\partial}{\partial\sigma^*(\vec x)}\,,
\quad
\hat\gamma^\dagger(\vec x)=-i\frac{\partial}{\partial\sigma(\vec x)}\,.
\label{eq:B15}
\end{align}
This yields eqs.~\eqref{eq:I2}--\eqref{eq:I4}.
The term $-q^{(0)}$ arises from the
terms involving two field derivatives. 
A more detailed stepwise computation can be found in the appendix.

\indent From our construction we know the identity
\begin{equation}
\partial_t q^{(1)}[\psi_S]
=
-iH_Lq^{(1)}[\psi_S]
=
q^{(1)}[-iH_S\psi_S]\,.
\label{eq:B16}
\end{equation}
This central result may be verified by using the explicit form of $q^{(1)}$.
For every complex one-particle wave function $\psi_S(\vec x)$ we can
construct the corresponding real wave function $q^{(1)}[\psi_S(\vec x)]$
and corresponding one-particle probability distribution $w_1[\psi_S (\vec x)]=\bigl(q^{(1)}[\psi_S(\vec x)]\bigr)^2$.
Solutions of the one-particle Schr\"odinger equation \eqref{eq:76} for
$\psi_S(t,\vec x)$ map to corresponding solutions of the Liouville
equation for $q^{(1)}(t) = q^{(1)}[\psi_S(t,\vec x)]$. This provides an explicit
construction of probability distributions for classical fields whose time
evolution induces the time evolution of the quantum one-particle wave function
according to the Schr\"odinger equation with Hamiltonian $H_S$. This
holds for an arbitrary potential $V(\vec x)$. In consequence, these wave
functions display all characteristic quantum effects such as interference
or tunneling.

\indent In contrast to $q^{(0)}$, the one-particle wave function $q^{(1)}$ cannot
be written as a product of two factors where each involves only one of the
flavors. The corresponding one-particle probability distribution involves
correlations between the two flavors.

\subsection*{Conserved charges for classical probabilities}

\indent Let us investigate how the charge observable characterizes the classical probability distribution.
So far we have only discussed the action of the charge operator on the wave function $\psi(\varphi)$ in eq.~\eqref{eq:29}. 
We need to specify the action of $\hat Q$ on the factor $\psi^*(\chi)$.

\indent One can define two local charge operators by
\begin{align}
\hat Q_\pm(\vec x)
&=
\hat\varphi_2(\vec x)\hat\eta_1(\vec x)
-
\hat\varphi_1(\vec x)\hat\eta_2(\vec x)
\nonumber\\
&\quad
\pm\left(
\hat\chi_2(\vec x)\hat\beta_1(\vec x)
-
\hat\chi_1(\vec x)\hat\beta_2(\vec x)
\right)
\nonumber\\
&=
\varphi^*(\vec x)\frac{\partial}{\partial\varphi^*(\vec x)}
-
\varphi(\vec x)\frac{\partial}{\partial\varphi(\vec x)}
\nn\\
&\quad\pm
\left(
\chi^*(\vec x)\frac{\partial}{\partial\chi^*(\vec x)}
-
\chi(\vec x)\frac{\partial}{\partial\chi(\vec x)}
\right)\,,
\label{eq:A1}
\end{align}
where
\begin{align}
\hat Q_+(\vec x)
&=
2\Big(
\sigma_2(\vec x)\pi_1(\vec x)
-
\sigma_1(\vec x)\pi_2(\vec x) \Big)
\nn \\
&\quad+
\frac12\left(
\frac{\partial^2}{\partial\pi_2(\vec x)\partial\sigma_1(\vec x)}
-
\frac{\partial^2}{\partial\pi_1(\vec x)\partial\sigma_2(\vec x)}
\right)\,,
\label{eq:A2}
\end{align}
and
\begin{align}
\hat Q_-(\vec x)
&=
i\biggl(
\sigma_1(\vec x)\frac{\partial}{\partial\sigma_2(\vec x)}
-
\sigma_2(\vec x)\frac{\partial}{\partial\sigma_1(\vec x)}
\nn \\
&\quad+
\pi_1(\vec x)\frac{\partial}{\partial\pi_2(\vec x)}
-
\pi_2(\vec x)\frac{\partial}{\partial\pi_1(\vec x)}
\biggr)\,.
\label{eq:A3}
\end{align}
The (global) charges 
\begin{equation}
\hat Q_\pm=\int_{\vec x}\hat Q_\pm(\vec x)
\label{eq:A4}
\end{equation}
are both conserved,
\begin{equation}
[\hat Q_\pm,H_L]=0\,.
\label{eq:A5}
\end{equation}
The operator $\hat Q_+$ is real in the basis $q(\sigma,\pi)$ and therefore
well defined. The operator $\hat Q_-$ is the generator of rotations in the
$(\sigma,\pi)$-space. Due to the factor $i$ it does not operate in the
space of real $q$.
If $\tilde \psi$ obeys the constraint \eqref{eq:24}, $\hat Q_- \tilde \psi$ does not.
Nevertheless, all even powers of $\hat Q_-$ are real
operators which act within the space of real wave functions $q$. They
correspond to well defined conserved statistical observables.
Every eigenstate $\psi_Q$ of $\hat Q$ leads to an eigenstate $\tilde \psi$ of $\hat Q_-$ with eigenvalue $Q_-=0$. 
It therefore corresponds to a wave function $q$ and a probability distribution $w$ which is invariant under flavor rotations.
States with a fixed number of particles and antiparticles are eigenstates of $\hat Q$ with
eigenvalue $Q$. The corresponding wave function $\tilde\psi$ is an
eigenstate of $\hat Q_+$ and $\hat Q_-$ with eigenvalues $Q_-=0$, $Q_+=2Q$.

\indent In the other direction we may expand a general wave function
$\tilde\psi(\varphi,\chi)$ can be written as a double expansion in eigenstates of $\hat Q$,
\begin{equation}
\tilde\psi(\varphi,\chi)
=
\sum_{Q,Q'}\sum_{\alpha,\alpha'}
b_{Q\alpha,Q'\alpha'}
\psi_{Q\alpha}(\varphi)\psi_{Q'\alpha'}^*(\chi)\,,
\label{eq:A6}
\end{equation}
where $\alpha$, $\alpha'$ label independent eigenstates for a given value of
$Q$. With
\begin{equation}
\hat Q_\pm\tilde\psi(\varphi,\chi)
=
\sum_{Q,Q'}\sum_{\alpha,\alpha'}
(Q\pm Q')\,b_{Q\alpha,Q'\alpha'}
\psi_{Q\alpha}(\varphi)\psi_{Q'\alpha'}^*(\chi),
\label{eq:A7}
\end{equation}
the condition $\hat Q_-\tilde\psi=0$ implies
\begin{equation}
\hat Q_-\tilde\psi=0 \Rightarrow (Q-Q')\,b_{Q\alpha,Q'\alpha'}=0\,,
\label{eq:A8}
\end{equation}
or
\begin{equation}
b_{Q\alpha,Q'\alpha'}=\tilde b_{Q\alpha\alpha'}\,\delta(Q-Q')\,.
\label{eq:A9}
\end{equation}
In particular, for a pure state one has
\begin{equation}
b_{Q\alpha,Q'\alpha'}=
c_{Q\alpha}c^*_{Q'\alpha^\prime}\,,
\label{eq:A10}
\end{equation}
such that the condition $\hat Q_-\tilde\psi=0$ implies that $\psi(\varphi)$
is an eigenstate of $\hat Q$,
\begin{equation}
\psi_Q(\varphi)=\sum_\alpha c_\alpha\psi_{Q\alpha}(\varphi)\,.
\label{eq:A11}
\end{equation}

\indent One may impose for the real wave function $q(\sigma,\pi)$ the condition
\begin{equation}
\hat Q_-q(\sigma,\pi)=0\,.
\label{eq:A12}
\end{equation}
If $q$ corresponds to a factorized wave function \eqref{eq:25}
this condition projects to some eigenspace of $\hat Q$ with fixed $Q$.
Superpositions with different charges are eliminated by the condition \eqref{eq:A12}.

\indent The vacuum state has zero charges,
\begin{equation}
\hat Q_{\pm}\tilde\psi^{(0)}=0\,,\quad
\hat Q_{\pm}q^{(0)}=0\,.
\label{eq:A13}
\end{equation}
For the one-particle state with $Q=1$ one has
\begin{equation}
\hat Q_+\tilde\psi^{(1)}=2\tilde\psi^{(1)}\,,\quad
\hat Q_+q^{(1)}=2q^{(1)}\,.
\label{eq:A14}
\end{equation}
With the explicit expressions \eqref{eq:A2} \eqref{eq:A3} one verifies that the wave
functions \eqref{eq:83}, \eqref{eq:B5} obey the relations
\eqref{eq:A13}, \eqref{eq:A14}.
The combination of the conditions \eqref{eq:A12} and \eqref{eq:A14}, together
with the "pure-state condition" \eqref{eq:25}, restricts the wave function
to a state with fixed $Q=\pm1$. These conditions are preserved by the time
evolution and therefore define a closed subsystem.

\subsection*{Discrete particles}

\indent The particle-wave duality of quantum mechanics combines the propagation
of waves with the discreteness of particles. The probabilities for
finding a single particle at some position $\vec x$ are distributed
continuously. Nevertheless, in an array of detectors only one detector
will indicate the presence of the particle. For a description of particles
by a classical field theory the continuous wave aspect may perhaps seem
less surprising. Still, the wave aspect of quantum mechanics concerns the
continuous probability amplitudes or wave functions and is not directly
related to the continuous fields. For our description of the double-slit
experiment by classical probabilities a more detailed understanding of
the discrete particle aspect seems appropriate.
One may ask the simple question: what is a particle in a classical field
theory? Our answer could be summarized: 
a particle is a stable probabilistic subsystem in the space of probability
distributions for classical fields. This concept is actually rather close
to the concept of a single particle in a quantum field theory. A particle
is an excitation of the vacuum, the latter being a non-trivial probabilistic state. 

\indent The basic concept for the definition of the particle
subsystem is the discrete local particle number. The corresponding
conserved global particle number ensures stability. For simplicity, we use
a language with discrete space points $\vec x$.
For discrete $\vec x$ the local creation and annihilation operators obey the commutation relation
\begin{equation}
[a(\vec x),a^\dagger(\vec y)]
=
\delta(\vec x-\vec y),
\label{eq:86}
\end{equation}
where $\delta$ is now the discrete Kronecker symbol which equals one if
the discrete points $\vec x$ and $\vec y$ coincide, and vanishes
otherwise. The local particle number operator is given by
\begin{equation}
\hat n(\vec x)=a^\dagger(\vec x)a(\vec x)\,.
\label{eq:87}
\end{equation}
The eigenvalues of this operator are discrete integers. With $\hat n(\vec x)\,|0\rangle=0$ one has
\begin{align}
\hat n(\vec x)(a^\dagger(\vec y))^m\,|0\rangle &= [\hat n(\vec x), (a^\dagger(\vec y))^m]\,|0\rangle
\nn \\
&=m\delta(\vec x-\vec y)a^\dagger(\vec y)\,|0\rangle\,,
\label{eq:88}
\end{align}
and
\begin{equation}    
\hat{n} (\vec x) \left( b^\dagger (\vec{y}) \right)^{m^\prime}\,|0\rangle = 0\,.
\label{eq:89}
\end{equation}
Arbitrary wave functions $\psi(\varphi)$ can be constructed as linear combinations of basis states,
\begin{align}
\label{eq:132}
&\hat{\varphi}[m,m']
=
\prod_{\vec y}
\left(a^\dagger(\vec y)\right)^{m(\vec y)}
\left(b^\dagger(\vec y)\right)^{m'(\vec y)}
|0\rangle\,,
\\
&\psi(\varphi)
=
\sum_{[m,m']}
c[m,m']\,\hat{\varphi}[m,m']\,. \nn 
\end{align}
Here the sum extends over all configurations of integers $\{m(\vec y),m'(\vec y)\}$, that we denote by $[m,m']$.
Each basis vector denotes a state where  $m(\vec y)$ particles and $m'(\vec y)$ antiparticles are present on the site $\vec y$. 
The relations \eqref{eq:88}, \eqref{eq:89} specify these states as eigenstates of $\hat n(\vec x)$, with eigenvalues $n(\vec x) = m(\vec y  = \vec x)$ counting the number of particles present at the site $\vec x$.

\indent The local particle number $\hat n(\vec x)$ is a statistical observable which does not have a definite value for a particular field configuration $(\sigma,\pi)$. 
It assumes a fixed integer value only for wave functions for which a fixed number of particles is present at $\vec x$.
From $\hat n(\vec{x})$ we can construct the operator for the total particle number $\hat N_p = \sum_{x} \hat n(\vec{x})= \sum_n a_n^\dagger a_n$.
It commutes with $H^{(q)}$, cf. eq.~\eqref{eq:72}, and therefore also with $H_L$.
Wave functions with a fixed total particle number $N_p$ form subsystems which are closed under the evolution. 
The same holds for the total number of antiparticles $N_a$. 
One can therefore find subsystems with a fixed number of particles and antiparticles. 
They have a fixed charge $Q=N_p-N_a$. 
For pure state wave functions this extends directly to subsystems for the real wave function $q$ and probability distributions $w$ for classical fields. 
For the double-slit experiment we are interested in the single-particle subsystem $N_p=1$, $N_a=0$.

\subsection*{Discrete observables}

\indent We want to make predictions for discrete observables as detectors which by construction can either fire or not.
A detector at some position $\vec{x}$ which fires should indicate the presence of a particle at $\vec{x}$ or in a neighborhood of $\vec{x}$.
It should therefore be related to the local particle number $n(\vec{x})$, which is a statistical observable. 
We first propose a statistical detector observable with an associated operator whose expectation value can be associated with the probability to find a particle in a neighborhood of $\vec{x}$. 
Subsequently, we will discuss the issue of expectation values of statistical observables and their measurements in a more general approach.

\indent Let us define a detector observable $D(\vec x)$ for a detector localized at
$\vec x$. It can only take the values one - the detector fires - or
zero - the detector does not fire. One may associate to $\vec x$ a certain region
$I(\vec x)$ of points around $\vec x$. We require that $D(\vec x)$ takes
the value one for all probability distributions for which at least one
particle is present in $I(\vec x)$. This is realized if we associate to
the detector observable the operator
\begin{equation}
\hat D(\vec x)
=
\tilde \Theta\left( \sum_{\vec y \in I(\vec x)} \hat n (\vec y) \right)\,.
\label{eq:92}
\end{equation}
This operator takes the value one for all basis states for which $\hat n(\vec y)$ is larger than zero for at least one point $\vec y\in I(\vec x)$. 
We extend this to all linear combinations of basis states for which at least for one value $\vec y\in I(\vec x)$ one has a fixed value $n(\vec y)>0$. On the other side, for all probability distributions which have a fixed value $n(\vec y)=0$ for all $\vec y\in I(\vec x)$ the detector variable takes a fixed value zero.
Probability distributions for which neither the first nor the second condition holds are not characterized by a fixed value of $D(\vec x)$.
These are linear combinations of basis states for which not all $n(\vec y\in I(\vec x))$ vanish, with the property that for every $\vec y\in I(\vec x)$ there are non-zero coefficients for at least two states which differ in $m(\vec y)$.

\indent We can still evaluate the expectation value $\langle D(\vec x)\rangle$
for an arbitrary wave function $\psi(\varphi)$,
\begin{equation}
\langle D(\vec{x})\rangle
=
\int \mathcal{D}\varphi\,
\psi^\dagger(\varphi)
\hat{D}(\vec{x})
\psi(\varphi)\,.
\label{eq:A35A}
\end{equation}
For a pure quantum state this comes over to the expectation value computed from the associated probability distribution $w$. 
We will use this for a prediction of the probability $\bar p(\vec x)$ that for a given $\psi (\varphi)$ the detector at $\vec x$ fires,
i.e. that $D(\vec x)$ takes the value one for this $\psi(\varphi)$,
\begin{equation}
\bar p(\vec x)=\langle D(\vec x)\rangle\,.
\label{eq:93}
\end{equation}
The concept of probabilities for measurements of statistical observables is similar to the concept of probabilities for certain values of the temperature or pressure. 
While temperature and pressure are themselves only characterizing a probability distribution, they can be measured and one may encounter ensembles for which such measurement values are distributed around some mean value. 
Mean values and correlations for statistical observables are useful concepts in macrophysics. 
They play a crucial role for microphysics if particles are characterized by statistical observables. 
Still, the association of the mathematical value of $\langle D(\vec x)\rangle$, as computed from the wave function, to the mean value of measurements of $D(\vec x)$ in some ensemble involves a detailed discussion of the particular measurement process into which we will not enter in this note. 
We present some general considerations below and refer to ref.~\cite{CWPW, CWPW2} for a discussion for other statistical observables. 
Our assumption is that standard particle detectors realize the relation \eqref{eq:93}. 
A similar assumption is used for the measurement of particles in standard quantum field theories.

\indent We next turn to the case where the wave function is characterized by a single particle. 
The total particle number has the sharp value
\begin{equation}
N_P=\sum_{\vec x}n(\vec x)=1\,.
\label{eq:94}
\end{equation}
This property is conserved by the evolution. 
We consider a subsystem with no antiparticle, such that the number of antiparticles has the sharp value zero. 
The only basis functions which contribute to the wave function for the one-particle subsystem have $m(\vec y)=1$ for precisely one value of $\vec y=\vec x$, and $m(\vec y)=0$ for all other sites $\vec y \neq \vec x$. 
Furthermore, $m'(\vec y)=0$ for all $\vec y$. 
The non-vanishing coefficients $c[m,m^\prime]$ reduce to $c_{10}(\vec x)$, corresponding to $\psi_S(\vec x)$ in eq.~\eqref{eq:65}.
For the one-particle subsystem the operator for the detector observable $D(\vec x)$ takes a very simple form. 
It takes the value one for all basis functions $a^\dagger(\vec y)\,|0\rangle$ if $\vec y\in I(\vec x)$, and zero for all other basis functions. 
Transfering to an operation on the one-particle wave function $\psi_S(\vec x)$ one has
\begin{equation}
\hat D(\vec x)\psi_S(\vec y)
=
\begin{cases}
\psi_S(\vec y), & \vec y\in I(\vec x),\\
0, & \text{otherwise}.
\end{cases}
\label{eq:95}
\end{equation}
Correspondingly, the probability $\bar p(\vec x)$ that the detector fires is given by
\begin{align}
\bar p(\vec x)
&=
\langle D(\vec x)\rangle
=
\sum_{\vec y}
\psi_S^*(\vec y)\,\hat D(\vec x)\,\psi_S(\vec y)
\nonumber\\
&=
\sum_{\vec y\in I(\vec x)}
\psi_S^*(\vec y)\psi_S(\vec y)\,.
\label{eq:96}
\end{align}
This is precisely the quantum mechanical probability to find a particle
in the interval $I(\vec x)$. By definition of the detector observable the
only possible outcomes of a measurement are the discrete values one or zero.

\subsection*{Expectation values of statistical observables}

\indent We have argued that statistical observables play an important role for a classical statistical description of microphysics.
Some words about the conceptual and mathematical status of their expectation values and the relation to measurements seem in order.
We may start by the observation that the notion of microphysical observables and their measurement is an idealization.
For a classical statistical description of molecules in a droplet of water nobody will ever
measure the precise locations and velocities of the molecules. The
concept of particles in phase space and associated classical observables,
together with a probability distribution over phase space, rather serves
as a basis for the computation of expectation values of more macroscopic
observables which can indeed be measured. This issue becomes even more
pronounced if one wants to describe a quantum particle by more fundamental classical fields. 
Even if we would put spacetime on a lattice, say with lattice distance given by the Planck length, there would be no way to measure precisely the classical field at a given point. 
The general strategy uses the idealisation of microscopic observables
for the computation of macroscopic observables which can be measured.

\indent We emphasize that the macroscopic observables are typically statistical
observables which have no definite values for points in phase space or
values of classical fields. They involve and therefore (partially)
measure properties of the probabilistic information. For the example of
liquids a typical observable is the density of the liquid at a certain location. 
This density is directly given by the probability distribution for finding a single molecule at this position. 
More generally, macroscopic observables often measure correlation functions.
These are again statistical observables.

\indent This raises the question which type of microscopic observables are available for the computation of the macroscopic observables.
They should include both classical observables and statistical observables.
We will deal with a large class of ``linear local observables'' where local refers to locality in time.
For these linear observables the mathematical expectation value is defined for a given value of the wave function by employing an associated real and symmetric operator $\hat{A}$,
\begin{align}
\langle A(t)\rangle
&=
q^T(t)\hat{A}q(t)
\nonumber\\
&=
\int_{\sigma,\pi,\sigma',\pi'}
q(\sigma',\pi';t)\,
\hat{A}(\sigma',\pi',\sigma,\pi)\,
q(\sigma,\pi;t)\,.
\label{eq:137A}
\end{align}
Every eigenvalue of $\hat{A}$ is assumed to be a possible outcome of the measurement of the observable $A$.
The linear local observables include the classical observables for which $\hat A$ is diagonal, given by a function $A(\sigma,\pi)$ without involving derivatives. 
They also encompass a very large family of statistical observables for which $\hat A$ involves derivatives. 
All observables discussed so far belong to the large class of linear local observables. 
From the concept of microscopic linear local observables one can compute the expectation values of the macroscopic observables which can be measured. 
An example for such macroscopic observables are the detector observables which may involve a sum over a very large number of ``Planck cells'' in space.

\indent The class of linear observables is wide enough to be compatible with the notion of subsystems. 
Classical observables are often mapped to statistical observables for the subsystem \cite{CWPO, CWPW, CWPW2}. 
A restriction of the microscopic observables to classical observables is typically not compatible with the notion of subsystems. 
The inclusion of statistical observables also allows for the definition of more robust microscopic observables. 
Classical field configurations are rather unstable microscopic observables. 
For probability distributions with a certain dispersion for $\pi(\vec{x})$ a precise value of $\sigma(\vec{x})$ would survive only for time scales much shorter than a typical experimental time scale.
The ``fluctuating field'' $\varphi(\vec{x})$ includes already a broadening
due to a dispersion of the time derivatives of fields \cite{CWQFTCF}. 
It is therefore a more robust microscopic variable.

\indent For a given macroscopic measurement apparatus with known outcomes -- say
one or zero for a detector firing or not -- one has to ask if there exists a macroscopic observable based on the linear microscopic observables with the correct properties. 
In the other direction, for a given mathematical macroscopic observable one may ask if one can construct an apparatus for an ``ideal measurement'' of its expectation value.

\subsection*{Born rule for classical statistical}

\indent A particular simple class of linear local observables are ``two-level observables'' for which the only possible measurement values are one and zero.
For two-level observables the associated operator $\hat{A}$ should have only the eigenvalues one and zero. 
(More general observables with only two different eigenvalues of the operator $\hat{A}$ can be brought to this form by a shift and a multiplicative factor.)
The expectation value of a two-level observable defines a probability distribution.
We can associate $\langle A\rangle$ with the probability $p_1$ to find the value $A=1$,
and $p_0=(1-\langle A\rangle)$ with the probability to find $A=0$.
Both $p_0$ and $p_1$ are positive, with $p_0+p_1=1$.
(This concept of defining a probability distribution from expectation values can be extended to more general observables if one also includes the expectation values of powers of the observable.)
For two independent two-level observables $A_1$ and $A_2$ the associated operators commute, $[\hat A_1,\hat A_2]=0$. 
We can then form combined probabilities, as
\begin{equation}
p_{01}=(1-\langle A_1\rangle)\langle A_2\rangle\,,
\label{eq:137B}
\end{equation}
for the probability to find $A_1=0$ and $A_2=1$. 
This extends to an arbitrary number of independent two-level observables. 
Their expectation values define a probability distribution. 
From this probability distribution one computes the probabilities for possible measurement values of macroscopic observables by the standard laws for classical probabilities.

\indent The probabilities computed from the expectation values of two-level observables can be cast into the form of a Born rule for classical statistics.
This applies to all linear observables, both classical and statistical.
Consider first a single two-level observable.
By a unitary transformation we can map $q$ to a complex wave function $\psi$ such that the operator $\hat{A}_d$ is diagonal,

$$
\psi=Uq\,,
\qquad
\psi=
\begin{pmatrix}
\psi_1\\
\psi_0
\end{pmatrix}\,,
\qquad
\hat{A}_d=U\hat{A}U^\dagger\,.
$$

\begin{equation}
\hat{A}_d
\begin{pmatrix}
\psi_1\\
0
\end{pmatrix}
=
\begin{pmatrix}
\psi_1\\
0
\end{pmatrix}\,,
\qquad
\hat{A}_d
\begin{pmatrix}
0\\
\psi_0
\end{pmatrix}
=
0\,.
\label{eq:BR1}
\end{equation}
In general, $\psi$ will be a complex function.
The diagonalization of $\hat{A}$ requires a unitary matrix which needs not to be a real orthogonal matrix.
Since $q$ is real, the complex wave function obeys the constraint
\begin{equation}
\psi^*
=
U^*U^\dagger\psi\,.
\label{eq:BR2}
\end{equation}

\indent Let us write $\psi$ in the form
\begin{equation}
\psi
=
\psi_{+\alpha}+\psi_{-\beta}\,,
\label{eq:BR3}
\end{equation}
where the indices $\alpha$ and $\beta$ denote the components of $\psi_+$ and $\psi_-$.
The normalisation of $\psi$ implies
\begin{equation}
\int \mathcal{D}\alpha\,
|\psi_{+\alpha}|^2
+
\int \mathcal{D}\beta\,
|\psi_{-\beta}|^2
=
1\,.
\label{eq:BR4}
\end{equation}
From
\begin{equation}
\langle A\rangle
=
\int\mathcal{D}\alpha\,
|\psi_{+\alpha}|^2\,,
\label{eq:BR5}
\end{equation}
one infers the probability to have $A=1$ as
\begin{equation}
p_1
=
\int\mathcal{D}\alpha\,
|\psi_{+\alpha}|^2\,.
\label{eq:BR6}
\end{equation}
and similarly
\begin{equation}
p_0
=
\int\mathcal{D}\beta\,
|\psi_{-\beta}|^2\,.
\label{eq:BR7}
\end{equation}
This is the Born rule for a single two-level observable.
It generalizes directly to an ensemble of commuting two-level observables.
In this case $U$ is chosen in order to diagonalize simultaneously all associated operators.

\indent Beyond two-level observables one can often decompose an observable into a set of two-level observables.
For example, an observable with four possible measurement values (the four eigenvalues of the associated operator) can be composed from two two-level observables, using the linearity of the rule \eqref{eq:137A}.
More generally, for linear observables we employ the unitary transformation from $q$ to $\psi$ such that in the $\psi$-basis $\hat{A}_d$ is diagonal.
We write $\psi=\sum_\lambda\psi_{\lambda,\alpha_\lambda}$, with $\psi_\lambda$ an eigenvector to the eigenvalue $\lambda$ of $\hat{A}$.
The probability that $A$ has the value $\lambda$ is given by
\begin{equation}
p_\lambda=\int\mathcal{D}\alpha_\lambda\,
|\psi_{\lambda,\alpha_\lambda}|^2\,.
\label{eq:BR8}
\end{equation}
This reproduces correctly the expectation value of $A$.
We can define functions of the observable $A$ by associating to $f(A)$ the operator $f(\hat{A})$.
The probabilities \eqref{eq:BR8} account for the expectation values of all functions $f(A)$ by
\begin{equation}
\langle f(A)\rangle=\sum_\lambda f(\lambda)p_\lambda\,.
\label{eq:BR9}
\end{equation}

\indent The probabilities $p_\lambda$ in eq.~\eqref{eq:BR8} describe a general Born rule for linear observables.
Instead of starting with $q$ we can start from the wave function in an arbitrary basis, using a unitary transformation to diagonalize $\hat{A}$.
The Born rule is particularly simple if one has a basis where $\hat{A}$ is already diagonal.
One can then read off the probabilities from the components of the wave function, with appropriate sums over degeneracies $\alpha_\lambda$.
For classical observables the basis where $\hat{A}$ is diagonal is given by $q$.
The Born rule yields directly the classical probabilities for the possible measurement values of $A$.
So far the probabilities are ``mathematical probabilities'' which correspond to the expectation values \eqref{eq:137A}.
The question of how they translate to sequences of real measurements concerns the issue of ``ideal measurements''.

\subsection*{Measurements of statistical observables}

\indent For an actual measurement of a macroscopic observable one would like to determine the probabilities for its possible measurement values by a sequence of identical measurements. 
The question arises if these ``frequentist'' probabilities coincide with the ones computed from the expectation values \eqref{eq:137A}. 
For a measurement apparatus with given possible outcomes this is the question if an appropriate operator can represent this measurement apparatus as a linear local observable.

\indent The outcomes of real macroscopic measurements are not unique. 
They depend on the concrete realisation of the measurement apparatus. 
An ideal measurement of some linear local observable reproduces the mathematical expectation value of this observable by the appropriate average of measurement values in a series of ``identical measurements''. 
We will assume that the detectors in a real measurement can be represented by a linear local observable $D(\vec{x})$ and realise an ideal measurement. 
The question is then which precise class of $D(\vec{x})$, or associated operator $\hat D(\vec{x})$, corresponds to the apparatus used.

\indent The choice of $\hat{D}(\vec{x})$ should obey a number of requirements.
First, for all wave functions for which the number of particles in a suitable neighborhood of $\vec{x}$ takes a sharp value one or zero, the expectation value $\langle D(\vec{x})\rangle$ should correspond to this value.
This implies that such wave functions are eigenfunctions of $\hat{D}(\vec{x})$ with the corresponding eigenvalues.
Second, the only possible outcomes of the measurement are one or zero -- the detector firing or not.
This implies that $\hat{D}(\vec{x})$ should not have any eigenvalues different from one or zero, or
\begin{equation}
\hat{D}^{\,2}(\vec{x})=\hat{D}(\vec{x})\,.
\label{eq:137C}
\end{equation}
Third, one requires ``extensiveness'' if one joins two regions in space.
Consider two regions $R_1$ and $R_2$ in space with detector observables $D(\vec{x}_1)$ and $D(\vec{x}_2)$.
The probability to find no particle in $R_1\cup R_2$ multiplies the probability $1-p_1$ to find no particle in $R_1$ with the probability $1-p_2$ to find no particle in $R_2$.
The probability to find a particle in $R_1\cup R_2$ is then given by the complement
\begin{equation}
p_{1,2}=1-(1-p_1)(1-p_2)=p_1+p_2-p_1p_2\,.
\label{eq:137D}
\end{equation}

\indent The operators (137) obey these requirements. 
They are therefore a good choice for representing an array of detectors. 
On the other hand, an array of particle detectors seems suitable for an ideal measurement of the statistical detector observable. 
The choice of a statistical observable for the description of the outcome of particle detectors seems crucial. 
We see no classical observable which takes fixed values for the field configurations $\sigma(\vec{x})$, $\pi(\vec{x})$ with the appropriate properties.

\indent This completes our classical probabilistic description of the double-slit experiment for the quantum particle. 
One chooses a classical field theory for which the potential $V(\vec x)$ in the field equations \eqref{eq:1}, \eqref{eq:2}, \eqref{eq:19A} realizes the double slit situation. 
One places a screen at $x_1=0$, realized typically by a diverging potential. 
One leaves open two slits centered at two values of $x_2$, and considers translation invariance in the coordinate $x_3$. 
An ensemble of detectors $D(x_1^{(D)},x_2,x_3)$ is placed at a distance $x_1^{(D)}>0$ from the screen. 
Starting with a one-particle wave function characterized by a wave packet with momentum $(p_1,0,0)$ centered at $x_1<0$, this wave packet evolves according to the Schr\"odinger equation \eqref{eq:76} which equals the one of the standard quantum particle in this potential. 
One can use the standard quantum mechanical computation of the expectation value of $\langle D(\vec x)\rangle$ in eq.~\eqref{eq:96}. 
It shows the typical interference effect of quantum mechanics. 
The detectors indicate the presence of the particle with the probability $\langle D(\vec x)\rangle$.

\subsection*{One-particle observables}

\indent For a single quantum particle moving in a potential characteristic observables are its position, momentum or angular momentum. 
One would like to describe these observables within the probabilistic classical field theory considered here. 
For this purpose we employ the detector observable $D(\vec x)$ which can take the possible measurement values one or zero. 
Its expectation value $\langle D(\vec x)\rangle$ is defined for an arbitrary wave function $q(t;\sigma,\pi)$ and therefore for an arbitrary probability distribution $w(t;\sigma,\pi)$. 
We assume that for an ideal measurement the probability $p(\vec x)$ to detect a particle at the detector position $\vec x$ is given by $\langle D(\vec x)\rangle$. 
For the one-particle state one has
\begin{equation}
\sum_{\vec x}\langle D(\vec x)\rangle
=
\sum_{\vec x}p(\vec x)
=
1\,.
\label{eq:Q01}
\end{equation}
Here the sum extends over the coarse lattice of detector positions which is supposed to cover the whole region of interest. 
The particle has to be somewhere in this region.

\indent The position observable $X$ is defined as
\begin{equation}
X=\sum_{\vec x}\vec x\,D(\vec x)\,.
\label{eq:Q02}
\end{equation}
For an eigenstate of $D(\vec x)$ only one detector fires at one particular site $\vec x_D$ with probability one. 
In this case the position observable takes the sharp value $X=\vec x_D$.
For a general one-particle state the position is not sharp. 
The expectation value,
\begin{equation}
\langle X\rangle
=
\sum_{\vec x}\vec x\,\langle D(\vec x)\rangle
=
\sum_{\vec x}\vec x\,p(\vec x)\,,
\label{eq:Q03}
\end{equation}
has the natural interpretation that each position $\vec x$ is weighted
with the probability $p(\vec x)$.
We can associate to $X$ the operator $\hat X$ acting on one-particle states
\begin{equation}
\hat X=\sum_{\vec x}\vec x\,\hat D(\vec x)\,.
\label{eq:Q04}
\end{equation}
The expectation value of $X$ obeys the quantum rule
\begin{align}
\langle \hat X\rangle
&=
\int_{\vec y}
\psi_S^*(\vec y)\,\hat X\,\psi_S(\vec y)
\nn\\
&=
\sum_{\vec x}\vec x
\int_{\vec y\in I(\vec x)}
\psi_S^*(\vec y)\psi_S(\vec y)
\nn\\
&=
\sum_{\vec x}
\vec x\,
\tilde\psi_S^*(\vec x)\tilde\psi_S(\vec x)\,.
\label{eq:Q05}
\end{align}
Here $\tilde\psi_S(\vec x)$ is a coarse grained wave function with
arguments on the detector-lattice. 

\indent For sufficiently smooth wave functions one can take the continuum limit,
\begin{equation}
\langle \hat X\rangle =\int_{\vec x}
\vec x\,
\psi_S^*(\vec x)\psi_S(\vec x)\,.
\label{eq:Q06}
\end{equation}
This is precisely the expectation value for quantum mechanics. 
Our discussion extends directly to observables which are arbitrary functions $f(X)$ of the position observable, with associated operator
\begin{equation}
\hat f(X)=\sum_{\vec x}
f(\vec x)\,\hat D(\vec x)\,.
\label{eq:Q07}
\end{equation}
The expectation value obeys
\begin{equation}
\langle \hat f(X)\rangle
=
\sum_{\vec x}
f(\vec x)\,p(\vec x)\,.
\label{eq:Q08}
\end{equation}
One finds the usual results of quantum mechanics for the dispersion by computing $\langle X^2\rangle-\langle X\rangle^2$.
In the following we take the continuum limit in space,
\begin{equation}
\langle f(X)\rangle=\int_{\vec x}
\psi_S^*(\vec x)f(\vec x)\psi_S(\vec x)\,.
\label{eq:Q09}
\end{equation}

\indent We can compute the mean velocity
\begin{equation}
\langle V(t)\rangle=\frac{1}{\varepsilon}
\left(
\langle X(t+\varepsilon)\rangle-\langle X(t)\rangle
\right)\,.
\label{eq:Q010}
\end{equation}
It involves the mean position at two neighboring times. 
Taking the time-continuum limit one has
\begin{align}
\langle V(t)\rangle
&=
\int_{\vec x}
\left(
\partial_t\psi_S^*(\vec x)\,
\vec x\,\psi_S(\vec x)
+
\psi_S^*(\vec x)\,
\vec x\,\partial_t\psi_S(\vec x)
\right)
\nn\\
&=
i\int_{\vec x}
\psi_S^*(\vec x)
\left(
H_S\vec x-\vec x H_S
\right)
\psi_S(\vec x)\,.
\label{eq:Q011}
\end{align}
This allows us to define a velocity observable $V(t)$ whose expectation value can be computed from the time-local probabilistic information encoded in the wave function $q^{(1)}(t;\sigma,\pi)$, and the associated one-particle probability distribution. 
Associating to $V$ the operator $\hat V$,
\begin{equation}
\hat V=i[H_S, X]\,,
\label{eq:Q012}
\end{equation}
one finds for every time $t$ the quantum rule
\begin{equation}
\langle V(t)\rangle =\int_{\vec x}
\psi_S^*(\vec x)\,
\hat V\,
\psi_S(\vec x)\,.
\label{eq:Q013}
\end{equation}
In the non-relativistic limit \eqref{eq:63} eq.~\eqref{eq:Q012} yields
\begin{equation}
\hat V=-\frac{i}{2m}
[\partial_x^2,\vec x]
=-\frac{i}{m}\partial_x\,.
\label{eq:Q014}
\end{equation}
Velocity and position cannot have simultaneously sharp values according to
\begin{equation}
[\hat V_k,\hat X_l]
=
-\frac{i}{m}\delta_{kl}\,.
\label{eq:Q015}
\end{equation}
This follows directly from Heisenberg's uncertainty relation using eqs.~\eqref{eq:Q09}, \eqref{eq:Q013}, \eqref{eq:Q014}. 
According to eq.~\eqref{eq:Q014} a sharp velocity requires a constant gradient of the wave function, which contradicts a sharp position.

\indent A momentum observable can be defined by detector observables firing at different times and positions,
\begin{equation}
P=mV\,,
\quad
\hat P=-i\partial_x\,.
\label{eq:Q016}
\end{equation}
This is close to experimental situations where one measures momentum as a function of particle trajectories, as in a tracking chamber.
Momentum and position operators have the standard form of quantum mechanics with commutator
\begin{equation}
[\hat P_k,\hat X_l]=-i\delta_{kl}\,.
\label{eq:BC17}
\end{equation}

\indent We observe that $\hat P=-i\partial_x$ is the generator of infinitesimal translations in space. 
This may be considered as a more profound definition of the momentum observable. 
Sharp values of momentum correspond to probability distributions which are periodic in space \cite{CWPW, CWPW2}. 
In this view the role of momentum as a statistical observable characterizing properties of the probability distribution is apparent. 
The generators of translations can be formulated on the level of the wave function $q$.
We express $q(\tilde\sigma,\tilde\pi)$ in terms of the fields
$\tilde\sigma(\vec x)$, $\tilde\pi(\vec x)$ defined by eq.~\eqref{eq:4A}. 
The generators of translations are then given by $(\partial_k=\partial/\partial x_k)$
\begin{align}
\hat P_k
=
-i\int_{\vec x}
\Bigg(
&
\partial_k\tilde\sigma(\vec x)
\frac{\partial}{\partial\tilde\sigma(\vec x)}
+
\partial_k\tilde\sigma^*(\vec x)
\frac{\partial}{\partial\tilde\sigma^*(\vec x)}
\notag\\
&+
\partial_k\tilde\pi(\vec x)
\frac{\partial}{\partial\tilde\pi(\vec x)}
+
\partial_k\tilde\pi^*(\vec x)
\frac{\partial}{\partial\tilde\pi^*(\vec x)}
\Bigg)\,.
\label{eq:Q018}
\end{align}
They involve field-derivatives and are therefore statistical observables.
(Only even powers of $\hat P$ are compatible with a real wave function $q$.) 
Since the system decomposes into a quantum field theory for the field $\varphi(\vec x)$ and its time-reflected part for the mirror field $\chi(\vec x)$, we can define the generator of translations for the pure state wave function $\psi(\varphi)$ in eq.~\eqref{eq:25},
\begin{equation}
\hat P_k^{(q)}\psi
=
-i
\int_{\vec x}
\left(
\partial_k\varphi(\vec x)
\frac{\partial}{\partial\varphi(\vec x)}
+
\partial_k\varphi^*(\vec x)
\frac{\partial}{\partial\varphi^*(\vec x)}
\right)
\psi\,.
\label{eq:Q019}
\end{equation}
The one-particle momentum obtains from $\hat P^{(q)}$ by projection to the one-particle state.

\indent Angular momentum can be defined similarly as the generator for space-rotations. 
For the one-particle subsystem the associated operator is the angular momentum operator of quantum mechanics.

\indent To sum up, typical one-particle observables are functions of position, momentum or angular momentum. 
They have associated operators which permit to compute their expectation values from the wave function $\psi_S$ of the one-particle subsystem. 
These operators are the usual ones of quantum mechanics.
The one-particle observables are statistical observables of the classical field theory. 
They typically obtain as the one-particle projection from statistical observables for the classical field theory. 
The quantum rule for the expectation values for one-particle observables follows from the expectation values of statistical observables in the classical field theory without invoking additional ``quantum axioms''. 
The relation between expectation values and mean values of measurements assumes the notion of ``ideal measurements'' \cite{CWPW2}.

\subsection*{Conclusions}

\indent We have presented a classical probabilistic description for a quantum
particle in an arbitrary potential. This includes all known ``quantum
mysteries'' as the interference in the double-slit experiment, tunneling
through a potential wall or the discrete spectrum of the hydrogen atom.
The key concepts for this embedding of a quantum system in a classical
probabilistic system are:
(i) the classical wave function encoding the probabilistic information,
(ii) statistical observables, and
(iii) subsystems for particle excitations of a vacuum. 
Many aspects are similar to quantum field theory. 
We start, however, with a positive probability distribution for field configurations.
The classical wave function is simply the root of the probability distribution.
Formulating the evolution in discrete time steps as a probabilistic cellular automaton \cite{ULA, VNEUA, ZUS, WOL, THO} we 
can cast this into a functional integral \cite{CWQFTCF} with an Euclidean classical action (see also ref.~\cite{MSR, JAN, DOM, GOZ1, GRT, BER}).
This functional integral describes the overall probability distribution for events at all times.

\indent Similar to the situation in quantum field theory, a detailed description
of the measurement process is a complex issue. It may need a more profound
investigation. For the moment we may consider our association
\eqref{eq:93} for the probabilities of outcomes of measurements of the
statistical detector observable with its expectation value computed from
the wave function as an assumption for an ideal measurement.
Similar to quantum field theory we also assume implicitly that the detector observable discussed
here corresponds to an ideal measurement with real particle detectors.
The key issue concerns the mean values of statistical observables which can only take discrete sharp measurement values.
For ideal measurements they are associated to expectation values for corresponding operators \cite{CWPW, CWPW2}.
This relation is assumed on the level of the classical probabilistic theory and can be taken as a definition of an ideal measurement for those statistical observables.
From there the quantum rule for expectation values of one-particle observables follows without invoking additional ``quantum axioms''.

\indent An extension to several quantum particles in a potential is straightforward. 
As long as the classical field equation remains linear our setting can describe a wide variety of situations, for example particles in the presence of arbitrary (background) fields.
One can define suitable correlations between properties of two spatially separated particles. 
They are given by the usual quantum correlations and can violate Bell's inequalities.

\indent What is not covered are additional interactions between the quantum particles, as the interaction between electrons in the Coulomb potential of an atom. 
In our context this will require a classical field theory with interactions. 
For a non-linear field equation the simple factorization into a quantum field and a mirror field does no longer work for the fields $\varphi$ and $\chi$. 
It is an interesting question if one can find modified fields $\varphi'$ and $\chi'$, possibly related to $\varphi$ and $\chi$ by non-linear field transformations, for which the factorization continues to work.
This may be related to the task to find renormalized fields for which the correlation functions remain finite in the limit where an ultraviolet cutoff is taken to infinity.

\indent We also have described only bosons. 
An extension to fermions may employ the equivalence between fermionic quantum field theories and generalized classical Ising models \cite{CWFI, CWFQFT, CWCWF}.
The discrete occupation numbers for fermions or associated values of Ising spins provide for a very direct implementation of discrete observables.
It is not clear, however, which fermionic quantum field theories can be constructed from probabilistic classical generalized Ising models. 
It may well be that the assumption of embedding a quantum field theory into a classical probabilistic theory turns out to become a selection criterion for a fundamental quantum field theory.

\indent \textit{Acknowledgement:} The author thanks J. Berges, R. Percacci and W. Zurek for stimulating discussions.

\appendix
\subsection*{Appendix: Computation of one-particle classical wave function}

\indent For a stepwise computation of $q^{(1)}$ we start from eq.~\eqref{eq:65} with 
\begin{equation}
a^\dagger(\vec x)
=
\frac1{\sqrt2}
\left[
H_S^{\frac{1}{2}}\varphi^*(\vec x)
-
H_S^{-\frac{1}{2}}
\frac{\partial}{\partial\varphi(\vec x)}
\right]\,.
\label{eq:84}
\end{equation}
This yields the one-particle wave function \eqref{eq:65} as
\begin{equation}
\psi^{(1)}
=
\sqrt2\int_{\vec x}
\psi_S(\vec x)
H_S^{\frac{1}{2}}\varphi^*(\vec x)\,|0\rangle\,.
\label{eq:85}
\end{equation}

\indent The product reads \eqref{eq:25} reads
\begin{equation}
\tilde\psi^{(1)}
=
\int_{\vec x,\vec y}
H_S^{\frac{1}{2}} (\vec x)
\psi_S(\vec x)
\left( H_S^{\frac{1}{2}} (\vec y)\psi_S(\vec y) \right)^\ast
A(\vec x,\vec y)\tilde\psi^{(0)}\,,
\label{eq:B1}
\end{equation}
where $A(\vec x,\vec y)$ can be written in terms of the complex fields
$\varphi(\vec x)$ and $\chi(\vec y)$,
\begin{equation}
A(\vec x,\vec y)=2\varphi^\ast(\vec x)\chi(\vec y)\,.
\label{eq:B2}
\end{equation}
In this form we can evaluate the one-particle wave function in an
arbitrary basis. For example, in the basis $\tilde\psi(\sigma,\zeta)$
one has
\begin{align}
A(\vec x,\vec y)
=&
\sigma_1(\vec x)\sigma_1(\vec y)
+\sigma_2(\vec x)\sigma_2(\vec y) \nn \\
&
-\frac14\left(
\zeta_1(\vec x)\zeta_1(\vec y)
+\zeta_2(\vec x)\zeta_2(\vec y)
\right)
\nonumber\\
&
+\frac12\Big(
\zeta_1(\vec x)\sigma_1(\vec y)
+\zeta_2(\vec x)\sigma_2(\vec y)
\nonumber \\
&\qquad-\zeta_1(\vec y)\sigma_1(\vec x)
-\zeta_2(\vec y)\sigma_2(\vec x)
\Big)
\nonumber\\
&
+i\Bigg[
\sigma_1(\vec x)\sigma_2(\vec y)
-\sigma_2(\vec x)\sigma_1(\vec y)
\nonumber \\
&\qquad-\frac14\left(
\zeta_1(\vec x)\zeta_2(\vec y)
-\zeta_2(\vec x)\zeta_1(\vec y)
\right)
\nonumber\\
&\qquad+\frac12\Big(
\zeta_1(\vec x)\sigma_2(\vec y)
-\zeta_2(\vec x)\sigma_1(\vec y)
\nonumber \\
&\qquad+\zeta_1(\vec y)\sigma_2(\vec x)
-\zeta_2(\vec y)\sigma_1(\vec x)
\Big)
\Bigg]\,.
\label{eq:B3}
\end{align}
For the computation of the real wave function $q^{(1)} (\sigma, \pi)$ one replaces
\begin{align}
\zeta_j(\vec x)\tilde\psi^{(0)}
&\to
i\frac{\partial}{\partial\pi_j(\vec x)}\tilde q^{(0)}
=
-2i\,B^{-1/2}(\vec x)\pi_j(\vec x)\tilde q^{(0)},
\nonumber\\
\zeta_j(\vec x)\zeta_k(\vec y)
&\to
\Big\{
-4B^{-1/2}(\vec x)B^{-1/2}(\vec y)\pi_j(\vec x)\pi_k(\vec y)
\nn \\
&\qquad\qquad+2B^{-1/2}(\vec x)\delta_{jk}\delta(\vec x-\vec y)
\Big\}\tilde q^{(0)}.
\label{eq:B4}
\end{align}
This yields
\begin{align}
q^{(1)} &= \Bigg[ 
\int_{\vec x, \vec y} \Big\{ \tilde{\psi}_S (\vec x) \tilde{\psi}_S^\ast (\vec y)
\Big(\tilde{A}_R (\vec x, \vec y) 
\nn \\
&\hspace{3.5cm}+ i \tilde{A}_I (\vec x, \vec y) \Big)
\Big\}-1
\Bigg]q^{(0)}
\nn \\
&=
\Bigg[
\int_{\vec x,\vec y}
\Big\{
\operatorname{Re}\bigl(\tilde\psi_S(\vec x)\tilde\psi_S^*(\vec y)\bigr)
\hat A_R(\vec x,\vec y)
\nn \\
&\qquad\quad-
\operatorname{Im}\bigl(\tilde\psi_S(\vec x)\tilde\psi_S^*(\vec y)\bigr)
\hat A_I(\vec x,\vec y)
\Big\}
-1
\Bigg]q^{(0)}\,,
\label{eq:B5}
\end{align}
where
\begin{equation}
\tilde\psi_S(\vec x)=B^{1/4}(\vec x)\psi_S(\vec x)\,.
\label{eq:B6}
\end{equation}
One has
\begin{align}
\tilde A_R(\vec x,\vec y)
=
&\sigma_j(\vec x)\sigma_j(\vec y)
+
\bar\pi_j(\vec x)\bar\pi_j(\vec y)
\nn \\
&-\left(
\sigma_j(\vec x)\epsilon_{jk}\bar\pi_k(\vec y)
+
\sigma_j(\vec y)\epsilon_{jk}\bar\pi_k(\vec x)
\right)\,,
\label{eq:B7}
\end{align}
and
\begin{align}
\hat A_I(\vec x,\vec y)
=
&\sigma_j(\vec x)\bar\pi_j(\vec y)
-\sigma_j(\vec y)\bar\pi_j(\vec x)
\nn \\
&+
\sigma_j(\vec x)\epsilon_{jk}\sigma_k(\vec y)
+
\bar\pi_j(\vec x)\epsilon_{jk}\bar\pi_k(\vec y)\,,
\label{eq:B8}
\end{align}
with
\begin{equation}
\bar\pi_j(\vec x)=B^{-\frac{1}{2}}(\vec x)\pi_j(\vec x)\,,
\label{eq:B9}
\end{equation}
and $\epsilon_{12}=-\epsilon_{21}=1$, $\epsilon_{11}=\epsilon_{22}=0$.
Under $SO(2)$ flavor rotations $\epsilon_{jk}$ is invariant, and one
finds that $q^{(1)}$ is flavor-rotation invariant. 
Eqs.~\eqref{eq:B5}--\eqref{eq:B9} are equivalent to eqs.~\eqref{eq:I2}-\eqref{eq:I4}.

\nocite{*}
\bibliography{refs}

\end{document}